\documentclass[useAMS,usenatbib]{mn2e}
\usepackage{graphicx}

\newcommand{\msun}{${\rm M_{\sun}}$}

\def\ltsima{$\; \buildrel < \over \sim \;$}
\def\simlt{\lower.5ex\hbox{\ltsima}}
\def\gtsima{$\; \buildrel > \over \sim \;$}
\def\simgt{\lower.5ex\hbox{\gtsima}}
\def\kms{{\rm\,km\,s^{-1}}}

\def\kpc{{\rm\,kpc}}

\def\msun{{\rm\,M_\odot}}

\newcommand{\fmmm}[1]{\mbox{$#1$}}
\newcommand{\scnd}{\mbox{\fmmm{''}\hskip-0.3em .}}

\makeatletter
\newcommand\ion[2]{#1$\;${\small\rmfamily\@Roman{#2}}\relax}%
\makeatother

\def\deg{^\circ}

\def\s{\ifmmode \widetilde \else \~\fi}
\def\={\overline}

\def\spose#1{\hbox to 0pt{#1\hss}}

\def\lta{\mathrel{\spose{\lower 3pt\hbox{$\mathchar"218$}}
     \raise 2.0pt\hbox{$\mathchar"13C$}}}
\def\gta{\mathrel{\spose{\lower 3pt\hbox{$\mathchar"218$}}
     \raise 2.0pt\hbox{$\mathchar"13E$}}}
\def\Dt{\spose{\raise 1.5ex\hbox{\hskip3pt$\mathchar"201$}}}    
\def\dt{\spose{\raise 1.0ex\hbox{\hskip2pt$\mathchar"201$}}}    

\def\dotsfill{\leaders\hbox to 1em{\hss.\hss}\hfill}

\def\Gyr{{\rm\,Gyr}}

\loadboldmathitalic 
\title[The distribution of stars in the Milky Way analogue NGC~891]
{An HST/ACS investigation of the spatial and chemical structure and sub-structure of NGC~891, a Milky Way analogue$^1$}
\author[R. Ibata, M. Mouhcine, M. Rejkuba]
{Rodrigo Ibata$^{2}$, Mustapha Mouhcine$^{3}$ and Marina Rejkuba$^{4}$\\
$^{2}$Observatoire Astronomique de Strasbourg (UMR 7550),
      11, rue de l'Universit\'e, 67000 Strasbourg, France\\
$^{3}$Astrophysics Research Institute, Liverpool John 
      Moores University, Twelve Quays House, Egerton 
      Wharf, Birkenhead, CH41 1LD, UK\\
$^{4}$ESO, Karl-Schwarzschild-Strasse 2, 
      D-85748 Garching, Germany}

\date{\today}
\begin{document} 
\maketitle 
\begin{abstract}
We present a structural analysis of NGC~891, an edge-on galaxy that has long been considered to be an analogue of the Milky Way. Using starcounts derived from deep HST/ACS images, we detect the presence of a thick disk component in this galaxy with vertical scale height $h_Z=1.44\pm0.03\kpc$ and radial scale length $h_R=4.8\pm0.1\kpc$, only slightly longer than that of the thin disk. A stellar spheroid with a de Vaucouleurs-like profile is detected from a radial distance of $r \sim 0.5\kpc$ to the edge of the survey at $r \sim 25\kpc$; the structure appears to become more flattened with distance, reaching $q = 0.50$ in the outermost halo region probed. The halo inside of $r \sim 15\kpc$ is moderately metal-rich (median ${\rm [Fe/H] \sim -1.1}$) and approximately uniform in median metallicity. Beyond that distance a modest chemical gradient is detected, with the median reaching ${\rm [Fe/H] \sim -1.3}$ at $r \sim 20\kpc$. We find evidence for subtle, but very significant, small-scale variations in the median colour and density over the halo survey area. We argue that the colour variations are unlikely to be due to internal extinction or foreground extinction, and reflect instead variations in the stellar metallicity. Their presence suggests a startling conclusion: that the halo of this galaxy is composed of a large number of incompletely-mixed sub-populations, testifying to its origin in a deluge of small accretions.
\end{abstract}

\section{Introduction}

\footnotetext[1]{This work was based on observations with the NASA/ESA Hubble Space Telescope, obtained at the Space Telescope Science Institute, which is operated by the Association of Universities for Research in Astronomy, Inc.,under NASA contract NAS 5-26555. This publication also makes use of  data products from the  Two Micron
All Sky Survey, which is a  joint project of the University of Massachusetts and  the Infrared  Processing  and Analysis  Center/California Institute  of Technology, funded by the  National Aeronautics and Space Administration and
the National Science Foundation.}

Remarkable progress has been made in recent years in understanding galaxy formation and evolution. High redshift observations have revealed the star formation history of the universe and the evolution of galaxy morphology \citep[see, e.g.,][]{Bell:2005p5060, Ryan:2008p5049}. However, the nature of look-back observations does not allow one to study the evolution of individual galaxies, and low-mass or small-scale structures remain out of reach in all but the nearest galaxies. Hence observations in nearby galaxies are required to complement the samples at cosmological distances to answer such fundamental questions as: which building blocks of high redshift galaxies end up in what type of local galaxy?; how much of the stellar content of different galactic morphological components is formed in situ and how much is accreted?

In recent years, it has been increasingly recognised that many of the clues to the fundamental problem of galaxy formation are preserved in stars in the outskirts of galaxies, offering the unique opportunity to address these questions \citep{Freeman:2002p3007}. Hierarchical formation models suggest that galaxy outskirts form by accretion of minor satellites, predominantly at early epochs when large disk galaxies were assembling for the first time. The size, metallicity, and amount of sub-structure in the faint outskirts of present-day galaxies are therefore directly related to issues such as the small scale properties of the primordial power spectrum of density fluctuations and the suppression of star formation in small halos \citep{Springel:2005p3530}.

The Milky Way and the Andromeda galaxy (M31) have played a pivotal role in studies of the faint outer regions of galaxies over many decades \citep{Freeman:2002p3007}. Indeed, almost all we know about the properties of the faint outskirts of galaxies is based on observations of these two galaxies. In recent years, we have seen dramatic progress in large scale mapping of the Milky Way, with large surveys that are measuring (and will measure) photometry, kinematics, and chemical abundances of a large number of individual stars over a wide volume of the Galaxy (SDSS -- \citealt{Gunn:2006p3183, Ivezic:2008p4821}; RAVE -- \citealt{Steinmetz:2006p3534}; SDSS-II/SEGUE -- \citealt{Rockosi:2005p3527}; GAIA; Pan-STARRS). It is no exaggeration to say that the study of these data sets, to unravel the processes by which the Galaxy came into being, will be among the chief astrophysical goals for years to come. 

Although these Galactic surveys will significantly advance our understanding of the assembly process, it is unlikely that a sample of one galaxy will provide the definitive solution to the nature and origin of stellar populations in galaxies in general. A necessary step to fully exploit Galactic surveys and thereby establish a comprehensive picture for spiral galaxy formation is to determine whether the Galaxy is suitably typical by studying other giant spirals. It is precisely for this reason that the Andromeda galaxy has been surveyed in great detail in recent years both with wide-field photometry \citep{Ibata:2001p334, Ferguson:2002p300, Ibata:2007p160} and massive multi-object spectroscopy \citep{Ibata:2005p216, Chapman:2006p207, Kalirai:2006p2953, Guhathakurta:2006p2971}. These analyses show that the Milky Way and M31 are very different, with M31 displaying an abundance of signs of both ancient and recent merger activity \citep{Block:2006p2994, Ibata:2007p160}, while the Milky Way appears to have led an unusually quiet existence \citep{Hammer:2007p4615}. Significant effort is currently being devoted to interpret the cosmological implications of the substructures observed in the halos of nearby galaxies (e.g., \citealt{Font:2008p2999,Johnston:2008p9138}).

In the currently favored galaxy formation models, halos are populated by stars which are tidally-stripped from satellites as they fall into the host potential \citep{Abadi:2006p2934}. The structure, chemistry, and ubiquity of stellar halos are thus expected to reflect the details of the galaxy assembly process \citep{Bullock:2005p2996,Johnston:2008p9138}. The realization that hierarchical models of galaxy formation almost inevitably predict one to two orders of magnitude more dwarf galaxies than are actually seen around the Milky Way \citep{Moore:1999p3286,Klypin:1999p11293} has led to suggestions that most of the dwarfs formed in the early universe have dissolved into the halo \citep{Bullock:2000p2950}. 

Observationally however, the nature and the origin of galactic stellar halos remain elusive. Excitement about the formation of stellar halos has been sparked in recent years after the discoveries of disrupted dwarf galaxies and sub-structures in star samples in the Galaxy \citep{Ibata:1994p417, Yanny:2000p2948, Ibata:2003p280, Martin:2004p277} and in M31 \citep{Ibata:2001p334, Zucker:2004p3549, Ibata:2007p160}, reinforcing the concept that satellite accretion could be an important mechanism for building up galaxies. Studies of the inner parts of the M31 ``halo'' initially revealed dramatic differences between the halo populations of that galaxy and those of the Galaxy \citep{Durrell:2004p2983}. The Milky Way halo seems to be populated by old, metal-poor stars, while a few fields in the halo of M31 show a large population of intermediate age stars with a much higher overall metallicity \citep{Brown:2003p2966, Brown:2007p2995}. Red horizontal branch stars are also found as distant as 40 kpc from the nucleus of M31 \citep{Rich:2004p3522} suggesting that the stellar halo of M31 underwent considerably more enrichment than did the Galaxy halo. However, the conclusion that the M31 halo is globally younger and more metal-rich than that of the Milky Way cannot be drawn from extant data, since the fields in M31 in which the above results were obtained have been found to be significantly contaminated by various accretion events \citep{Ibata:2007p160}. Furthermore, spectroscopy of stars either selected by their kinematics to be halo members or sufficiently in the outskirts of the galaxy reveals the presence of a pressure-supported metal-poor population similar to the population that dominates the Galactic halo \citep{Chapman:2006p207, Kalirai:2006p2953}. The current observational evidence therefore demonstrates that halos are complex structures; and larger samples will be required to fully understand how they form and evolve.

To answer the fundamental question of how typical are the Local Group massive galaxies, and to constrain the assembly history of spirals, we need to probe beyond the Local Group, and we have therefore been studying the resolved extra-planar stellar populations of NGC~891 \citep{Mouhcine:2007p153, Rejkuba:2008p0000}. This galaxy is particularly interesting because (i) it is the nearest galaxy with similar morphology and mass to the Galaxy; (ii) previous studies have presented no indications of disturbance in the stellar component; and (iii) it is almost perfectly edge-on \citep[$i=89.8\deg\pm0.5\deg$ ---][]{Kregel:2005p11164}, essential to disentangle the halo and disk in the absence of kinematic information. 

\begin{figure}
\begin{center}
\includegraphics[angle=0,width=\hsize]{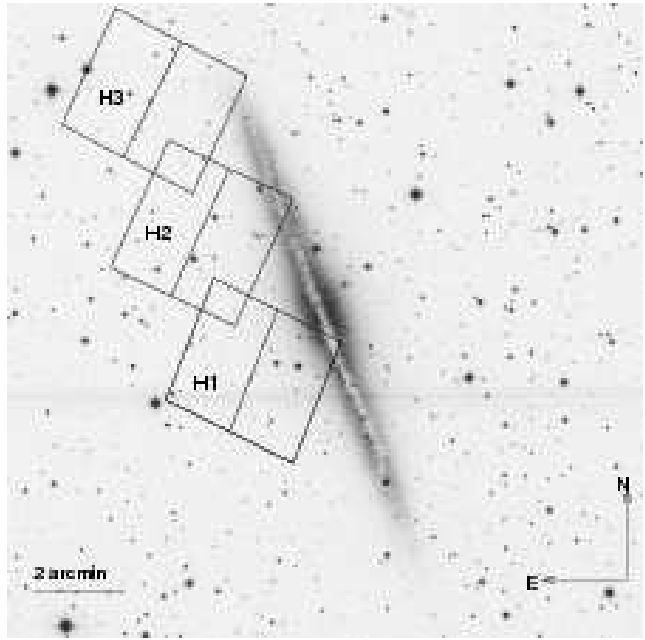}
\end{center}
\caption{The locations of the three ACS fields H1, H2 and H3, superimposed on photographic data from the Digitized Sky Survey.}
\end{figure}

One feature of NGC~891 that sets it apart from the Milky Way is that it possesses a substantial \ion{H}{1} halo \citep{Sancisi:1979p11158,Oosterloo:2007p5583}, which is due primarily to galactic fountain phenomena, although $\approx 10$\% may be of accretion origin \citep{Fraternali:2008p11174}. It appears that this gaseous component is linked to active on-going star-formation, and indeed the accretion rate matches closely the star-formation rate of $2.9\msun \, {\rm yr^{-1}}$ \citep{Fraternali:2008p11174}. An ionised gas ``halo'' component, almost certainly related to the star-formation activity, has been detected as an H$\alpha$ envelope surrounding the disk of the galaxy \citep{Dettmar:1990p11468,Rand:1990p11482,Rossa:2003p11269}. Star-formation in the Milky Way is known to be episodic, with numerous short-term peaks in the last $2\Gyr$ during which the star-formation rate reached $3$--$5$ times the steady rate \citep{delaFuenteMarcos:2004p11200}. Given the supernova-driven galactic fountain, it is likely that NGC~891 is currently undergoing such an episode. Since there is no reason to expect a causal connection between this halo gas and the halo or thick disk stellar populations, which in the Milky Way and in simulations are exclusively composed of ancient stars, we believe that NGC~891 can be examined as a plausible analogue of a galaxy formed in a similar manner to the Milky Way.

As described in detail in a companion paper to this contribution \citep[][hereafter Paper~II]{Rejkuba:2008p0000}, three fields in NGC~891 were observed with the Wide-Field Camera of the Advanced Camera for Surveys (ACS) on board the Hubble Space Telescope in cycle 11 (programme GO-9414). These three archival fields probe both the disk and thick disk of the galaxy over a wide radial range as well as the inner halo (see Fig.~1), and have slight overlaps which are very convenient as a means to ensure photometric consistency, and to check the reliability of the photometric uncertainties. With exposures of 7712~sec in both F606W (similar to broad-band $V$) and F814W (similar to broad-band $I$) the photometry reaches to a limiting magnitude of ${I\sim 29}$, approximately 3~magnitudes below the tip of the red giant branch (RGB). Paper~II gives a thorough description of the data and the data reduction process, describing how the images were combined, and how crowded-field point-source photometry was measured from the image stacks; we therefore refer the reader to Paper~II for these details.

As we showed in \citet[][hereafter Paper I]{Mouhcine:2007p153}, the properties of the extra-planar stars in NGC~891 questions the classical view of what a stellar halo should be. Our first analysis of the three contiguous ACS pointings showed that the metallicity distribution at $\sim10 \kpc$ perpendicular to the disk of the galaxy is dominated by stars with ${\rm [Fe/H]\sim -0.8}$. This is a full $\sim0.5$ dex more metal-rich than what is measured for the Galaxy halo stars at a comparable height, and suggests that the metal-poor stars observed in the Galaxy may not be the dominant population in galactic halos in general \citep{Mouhcine:2006p3310}. 

The purpose of the current contribution is to examine the spatial distribution of the stellar populations in more detail than in the previous two contributions in this series, and in particular to investigate the clumpiness of the halo population. The layout of this paper is as follows: in \S2 we begin our analysis by reviewing the large-scale spatial structure as derived from 2MASS $K$-band data; \S3 presents the spatial distribution of the ACS star-count data, focussing on the minor axis profile and the vertical profile of the outskirts of the disk. In \S4 we make a first attempt at uncovering small-scale sub-structures in the galaxy, and then proceed in \S5 to analyse the metallicity distribution over the survey area, and discuss the small-scale variations of the metallicity. The implications of these findings are discussed in \S6, and we draw the major conclusions from our study in \S7.

Throughout this paper we assume a distance of 9.73~Mpc (equivalent to a distance modulus of $(m-M)_0 = 29.94$, from Paper~I). We note that NGC~891 is located at relatively low Galactic latitude ($\ell=140.38^\circ$, $b=-17.42^\circ$), and therefore these fields suffer from significant (though not large) extinction from foreground dust: $E(B-V)= 0.065$ \citep{Schlegel:1998p5331}, corresponding approximately to $A_V=0.22$, $A_I=0.13$.

\begin{figure}
\begin{center}
\includegraphics[angle=-90,width=\hsize]{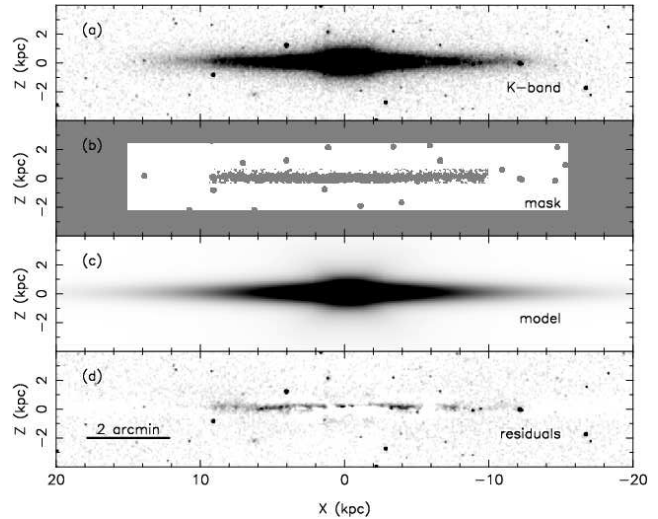}
\end{center}
\caption{The 2MASS $K$-band image of NGC~891 (upper panel), is fit within the area delimited by the mask (panel `b') with a two-component disk plus bulge model (panel `c'). The corresponding residuals are displayed in the bottom panel.}
\end{figure}

\section{Global model from $K$-band surface photometry}

To obtain an insight into the global large-scale structure of the galaxy, we first perform a fit to the 2MASS near-infrared data on this galaxy. The upper panel of Fig.~2 shows the $K$-band 2MASS Large Galaxy Atlas data \citep{Jarrett:2003p6828}, retrieved from the NASA/IPAC archive, rotated $68^\circ$  East of North to aid interpretation. In this $K$-band image the obscuring dust layer, which impedes easy analysis of the visible DSS image in Fig.~1, is very much attenuated. 

\citet{Xilouris:1998p6918} have previously made a detailed fit of this galaxy to $K$-band data (and other wavelengths), using a composite model consisting of a disk, a bulge and a dust-layer. For the purposes of our own study, we felt that their model could be slightly improved on. In particular, they did not try to fit the boxiness of the NGC~891 bulge, which is very marked as one can immediately perceive in Fig.~2. For reference, their $K$-band thin disk model has scale height $h_Z = 0.32\pm0.01\kpc$ and a scale length of $h_R = 3.93\pm0.1\kpc$, inclined at $\theta=89.6\deg \pm0.1\deg$ to the line of sight.

\begin{figure*}
\begin{center}
\hbox{
\hspace{1.3cm}
\includegraphics[angle=0,width=15.3cm]{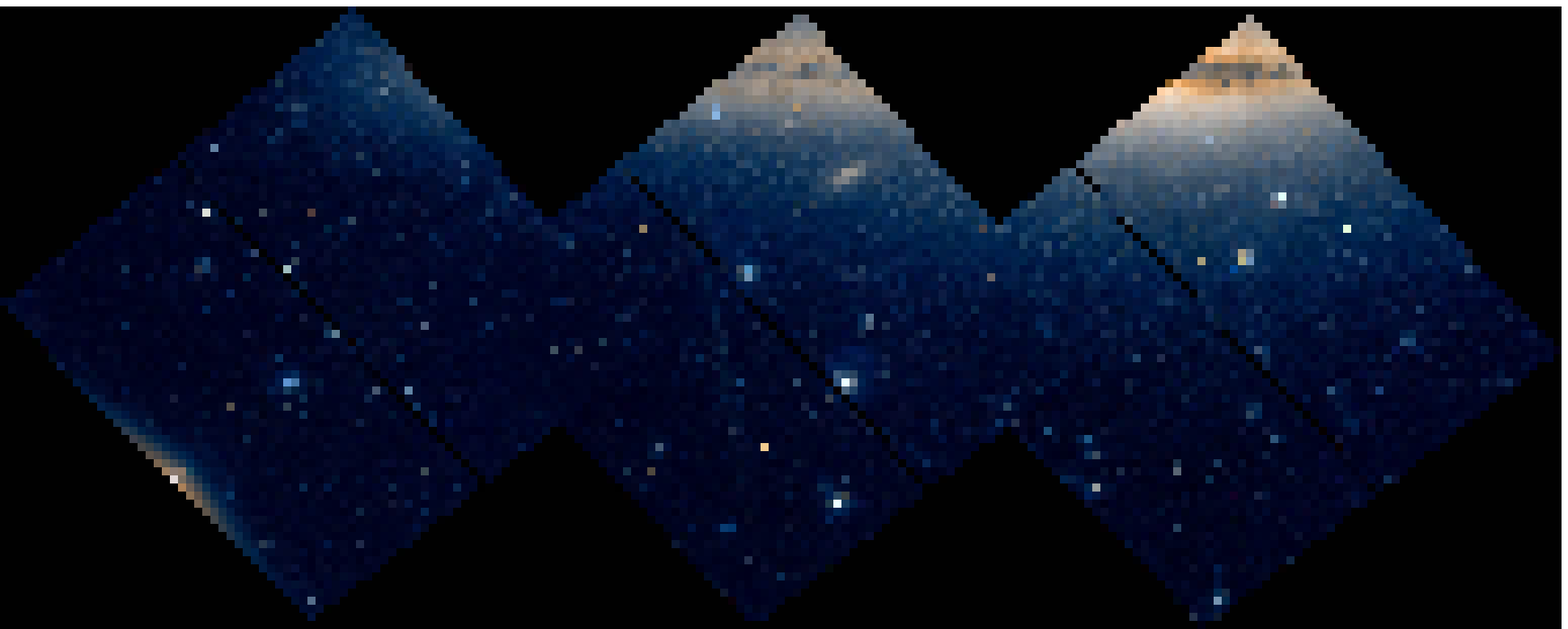}
}
\includegraphics[angle=-90,width=\hsize]{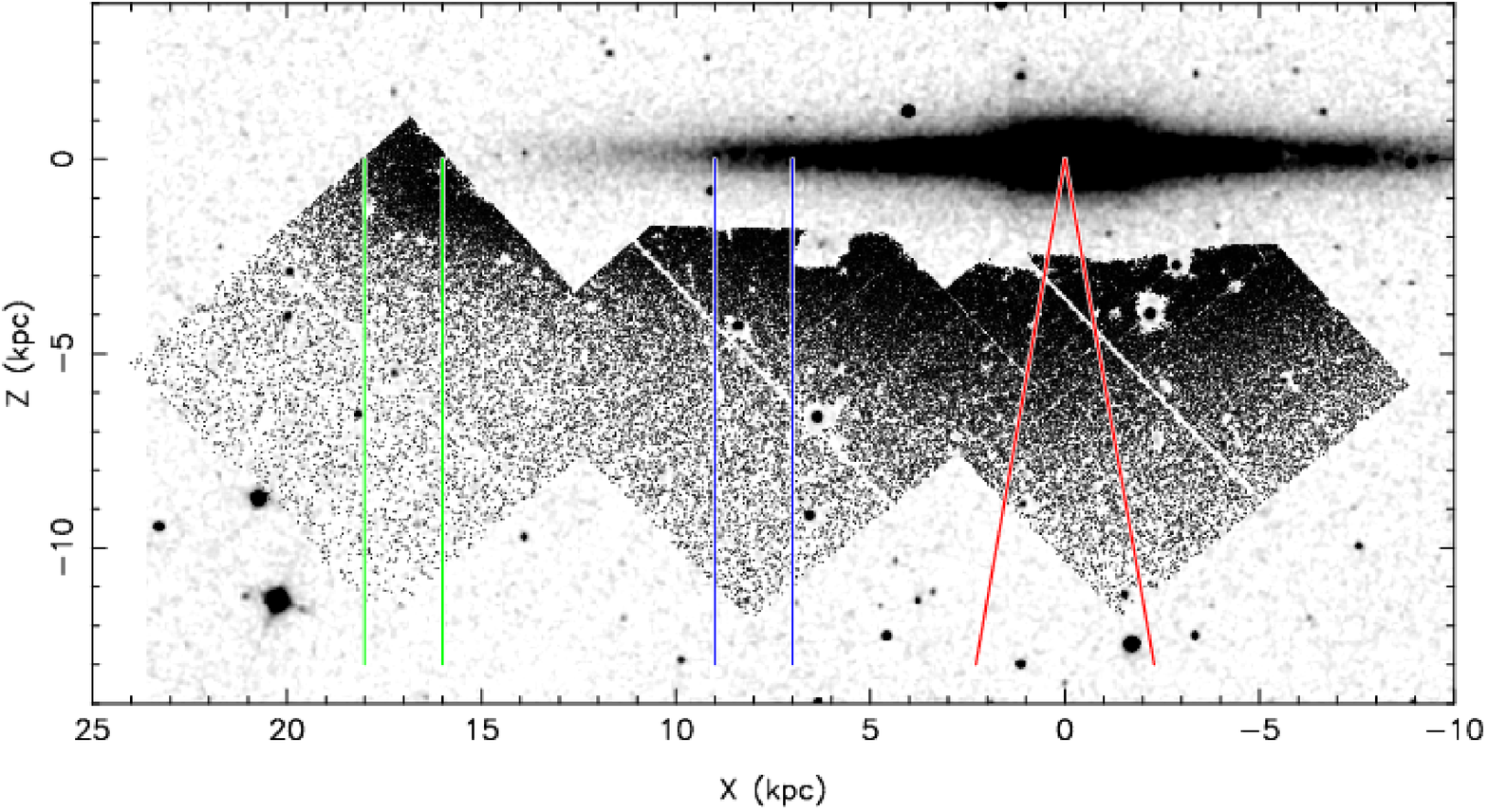}
\end{center}
\caption{The upper panel shows a colour mosaic of all three ACS fields, which highlights the crowding and extinction problems close to the galactic plane, as well as the diffraction spikes, bright star halos and reflected light artefacts. The spatial distribution of point-like sources is displayed in lower panel, superimposed on the $K$-band 2MASS data. The sample has been culled by keeping only reliable RGB candidates, as detailed in the text. The selection region for the minor axis sample (Fig.~5) is marked in red ($\pm 10\deg$ from the minor axis), that for the outer disk (Fig.~7a) is marked green ($16 < X < 18\kpc$), while that of the ``Solar-radius-like'' sample (Fig.~7b) is marked blue ($7 < X < 9\kpc$).}
\end{figure*}

We fitted the galaxy light distribution using the GALFIT software package v3.0 \citep{Peng:2002p3520}. The Large Galaxy Atlas images have $1\arcsec$ pixels ($0.047\kpc$), oversampling the original 2MASS frames by a factor of $2\times2$. For the purposes of determining the noise characteristics of the frame, we assumed (following \citealt{Maller:2005p6834}), that the image was derived from an average of 6 frames, each with 52~electrons read noise, and gain of 8.2 electrons per ADU. The disk was assumed to be a perfectly edge-on exponential (and hence $K_1(R/h_R)*(R/h_R)$ in projection, where $K_1$ is the modified Bessel-function), with the following free parameters: central surface brightness, scale length, scale height, position of centre and position angle. The bulge, however, was taken to be a Sersic model, with central surface brightness, effective radius $r_e$, Sersic exponent $n$, axis ratio $q$ (flattening), position of centre, and boxiness parameter $c$ (see \citealt{Peng:2002p3520} for a definition of this parameter). 

We limit the region to be fitted to $|X| < 15\kpc$, as this corresponds approximately to the region where the disk follows an exponential law \citep{vanderKruit:1981p5356}, and limit the fit in the vertical direction to $|Z| < 2.35\kpc$ ($50\arcsec$) to avoid including regions obviously devoid of galaxy signal in the 2MASS image (see Fig.~2). We masked out the stellar point sources in the frame (defined as 3-sigma deviations over the local background integrated over a 6.6~pixel aperture), blocking out circular regions of diameter three times the stellar FWHM. We also masked out the regions where the extinction is particularly strong and clumpy in the sub-area $|X| < 10\kpc$, $|Z| < 0.55\kpc$, which we identify as those pixels with ${\rm H-K > 0.4}$. The dark pixels in panel `b' of Fig.~2 show the areas that were masked out in this way.

Curiously, when we test the $K$-band structural model of \citet{Xilouris:1998p6918} on the $K$-band 2MASS data, including their dusty disk model, but assuming that the galaxy is perfectly edge-on, we do not find a particularly good fit. This slight variant of their model over-subtracts the thin disk, leaving behind a substantial excess of light at $|Z| > 0.25\kpc$. In principle, this cannot be due to the minor difference in choice of disk inclination: the slight offset of $0.4\deg$ between the edge-on model and the \citet{Xilouris:1998p6918} model would amount to a mere $\sim 0.14\kpc$ apparent offset from $Z=0$ for stars in the galactic plane located at $21\kpc$ (the truncation radius where the exponential profile drops precipitously \citealt{vanderKruit:1981p5356}) either in front or behind the galaxy. However, the exponential {\it model} they use (and that we use too) is infinite in extent, leading to a substantial contribution at high $|Z|$ even for very small disk inclinations; this is not physical however, as the real disk is truncated at $21\kpc$ \citep{vanderKruit:1981p5356}.

\begin{figure}
\begin{center}
\includegraphics[angle=0,width=\hsize]{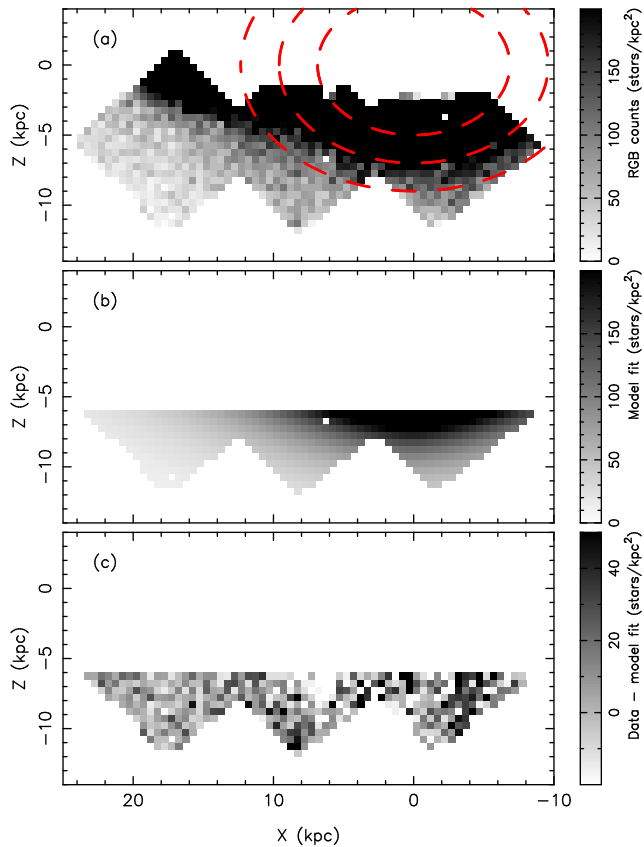}
\end{center}
\caption{Panel `a' shows the same data as Fig.~3, but with binned-up star-counts. The area under each of these $0.5\kpc\times0.5\kpc$ pixels is calculated from the ACS image weight maps. We have additionally masked out the haloes of bright stars with appropriately-sized circular regions. Ellipses of minor axis length 5, 7 and 9~kpc and axis ratio $0.73$ are shown to demonstrate the presence of an extended elliptical component. The middle panel displays a de Vaucouleurs model fit to the halo region at $Z < -6\kpc$, whose residuals are shown in the bottom panel.}
\end{figure}

We therefore interpret the residual excess light at $|Z| > 0.25\kpc$ from the \citet{Xilouris:1998p6918} model with $\theta=90\deg$ as a (slight) failing of that model due to the initial assumption that the disk is infinite. In the following we will also assume an infinite disk, but one that is perfectly edge-on; the outer regions are less problematic in this configuration, since their light is swamped by the much brighter inner regions of the disk. We feel that this assumption is justified by the V- and B-bands model fits of \citet{Xilouris:1998p6918}, which are consistent with the galaxy being perfectly edge-on.

For the reasons explained above, the \citet{Xilouris:1998p6918} model probably underestimates the disk scaleheight. However, since their fitted value of $h_Z=0.32\pm0.1\kpc$ is close to that of old stars in the Milky Way, it is informative to try to refit the galaxy with an additional thick-disk component to account for the excess light away from the galactic plane. Indeed, adding an additional disk to their model with a scale length constrained to be identical to their thin disk value ($h_R = 3.93\kpc$) gives an excellent fit with a resulting scale height of $h_Z = 0.76\pm0.01\kpc$.

However, the data do not require two stellar disks; fitting with a single disk and bulge model gives an excellent reduced $\chi^2$ ($=1.04$, a value calculated assuming only Poisson noise in the star-counts). The fitted disk scale length and scale height are found to be $h_R = 4.19\pm 0.01\kpc$ and $h_Z = 0.57\pm 0.01\kpc$, respectively. For comparison, the Galactic thin disk has $h_R = 2.6\kpc$ and $h_Z=0.3\kpc$ \citep{Juric:2008p4886}. Note, however, that we have masked-out most of the disk close to the galactic plane (Fig.~2b), which must bias our scale height to large values. We found the best-fitting parameters to be: S\'ersic index $n=4.33\pm0.10$ (close to a de Vaucouleurs profile), effective radius $r_e = 2.5\pm 0.1\kpc$, flattening $q = 0.73\pm0.01$, and boxiness $c = 2.1\pm0.1$. Panel `c' of Fig.~2 shows the resulting final model, and the residuals are displayed in the bottom panel. 

Although the residuals between the data and the model are quite small, there are large-scale patterns in the residual. These include the obvious thin structure close to the galactic plane, seen between $-10\kpc < X < 10\kpc$. Since this structure contains numerous point sources, which are clearly bright-star forming regions, it is likely a thin star-forming ring. The fact that it does not line up with the $Z=0$ axis and seems to undulate suggests an extra-planar influence in its formation. It is conceivable that this ring-like structure was formed in the interaction (or collision) with UGC~1809, recently claimed to be responsible for an extra-planar filament of \ion{H}{1} gas \citep{Oosterloo:2007p5583, Mapelli:2008p5516}. Note that this extra-planar population lies on the opposite side of the $Z=0$ plane to what would be expected from the \citet{Xilouris:1998p6918} inclination angle of $\theta=89.6\deg$, so it is not an artefact of our choice of assuming that the galaxy is perfectly edge-on. Another interesting aspect of the galaxy that is highlighted by the model fit in Fig.~2 is that the model over-subtracts the data beyond approximately $15\kpc$; this corresponds to the break radius, where the observed profile begins to drop much faster than the inner exponential disk \citep{vanderKruit:1981p5356}. The fit also confirms the ``boxy" appearance of the bulge, clearly seen in the upper panel of Fig.~2.

\section{Distribution of point-sources}

As can be seen in Fig.~1, there is significant overlap between the three ACS fields. Although we could have used the increased depth in the overlap regions to probe slightly deeper, we chose not to do so in order to maintain uniform sensitivity over the survey region. Furthermore, to avoid problems with variations of source completeness, we chose to keep in the overlap areas only those data from the higher numbered field (i.e. we kept the H3 data in the H2-H3 overlap, and the H2 data in the H1-H2 overlap). The bottom panel of Fig.~3 shows the distribution of point-sources detected with the DOLPHOT software package (a modified version of HSTphot, \citealt{Dolphin:2000p11500}) that have magnitudes $I > 25.8$, sharpness parameter less than 0.5 in both bands and crowding parameter less than 0.3. This choice of sharpness and crowding parameters is identical to that adopted in Paper~I. To be consistent with the colour-magnitude selection in Paper~I (see Fig.~3 of that contribution), we impose a faint limit $I < 27.4$ (for $V-I < 1.3$), then a linear increase to $I < 27.3$ (at ($V-I = 2.0$) and finally a linear increase to $I < 26.8$ (at $V-I=2.6$). As we show in Paper~II, this limit provides better than 50\% completeness even in the relatively crowded regions at $Z\sim -4\kpc$. The spatial distribution of the 87973 sources selected in this way is displayed in Fig.~3, superimposed on the 2MASS $K$-band image previously shown in Fig.~2.

\begin{figure}
\begin{center}
\includegraphics[angle=0,width=\hsize]{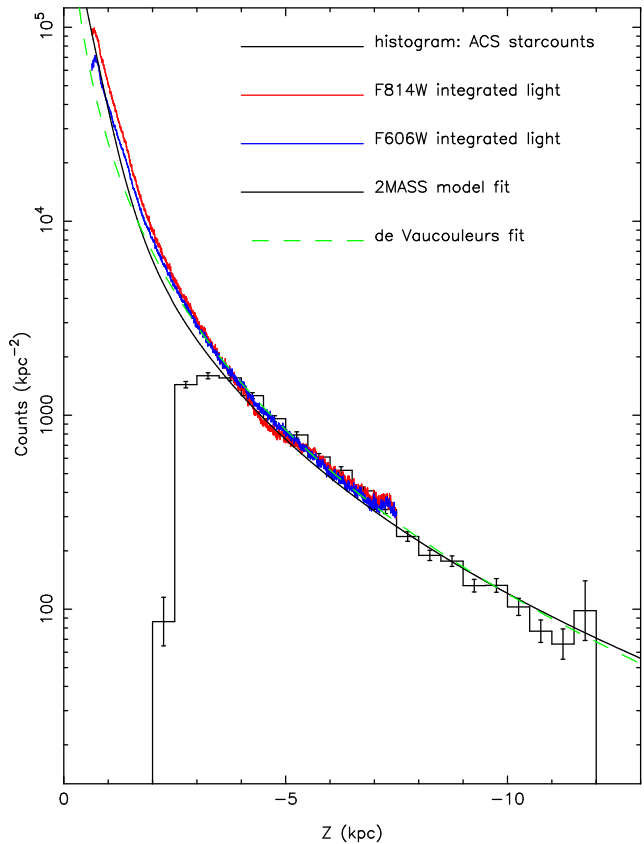}
\end{center}
\caption{The histogram shows the minor axis density distribution of the ACS star-counts, together with a fitted de Vaucouleurs model (green dashed line) which has effective radius of $r_e = 1.55\pm0.03$. The black line that is almost indistinguishable from the de Vaucouleurs profile for $Z < -2\kpc$ is the minor axis profile of the galaxy model previously fit in Fig.~2. The blue and red lines show, respectively, the minor axis surface brightness profiles derived directly from the HST/ACS images. In all cases a $\pm 10\deg$ wedge along the minor axis was selected for analysis (as indicated in the bottom panel of Fig.~3).}
\end{figure}

\begin{figure}
\begin{center}
\includegraphics[angle=-90,width=\hsize]{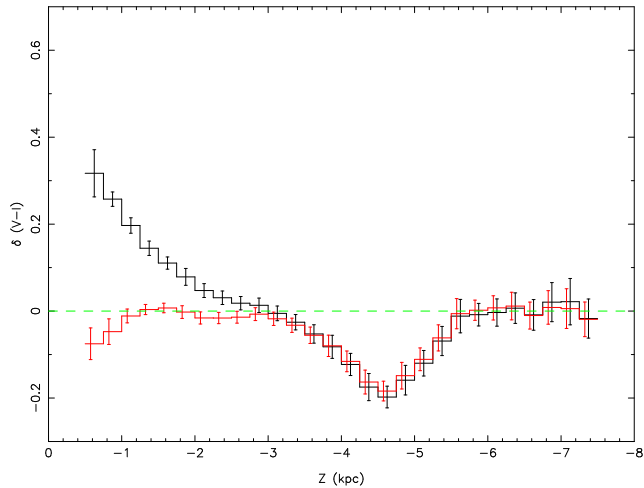}
\end{center}
\caption{The black line shows the distribution of (V-I) colour difference from integrated light, derived from a $\pm10\deg$ wedge on the minor axis. Given that the blue dip near $Z \sim -5.5\kpc$ is likely an artefact due to scattered light, the profile is remarkably flat for $Z < -2.5\kpc$, indicating that the stellar populations and extinction are very uniform at large distances from the galactic plane. Correcting for the reddening due to dust in the galaxy, as estimated from the \ion{H}{1} map of \citet{Oosterloo:2007p5583}, gives the red-line histogram. The error bars in both cases show the rms scatter in the median colour of the ACS pixels.}
\end{figure}

The irregular upper edge to the ACS point-sources in Fig.~3 (especially in fields H1 and H2) is due to the imposed crowding limit; beyond this area the accuracy of the photometry is compromised due to image blending. Also visible are the gaps between the ACS chips, as well as holes of low source density in the "haloes" around bright stars.

\subsection{The minor axis profiles}

A property of the stellar distribution that is immediately striking is the presence of a spheroidal distribution surrounding the bulge. This spheroidal population is seen more clearly if we bin the starcounts, as in Fig.~4, where we have overlaid ellipses of axis ratio 0.73 to guide the eye. The middle ellipse with minor axis length $7\kpc$ delimits quite well the ``edge" of this component. We stress that this is, of course, not a physical limit but simply an artefact of this particular greyscale representation. 

Is this ellipsoidal component the outer reaches of the bulge, or the inner parts of the halo? To answer this question we show in Fig.~5 the minor axis star-counts and the V and I-band surface brightness profiles. The star-counts profile (black histogram) was constructed by summing stars within a $\pm 10\deg$ triangular wedge around the minor axis (between the red lines in Fig.~3), accounting for lost detector area due to the various defects of the survey listed above. The blue and red lines in Fig.~5 show the surface brightness profiles from integrated light, measured in the F606W and F814W images respectively. These two profiles correspond to the median pixel value also in the same $\pm 10\deg$ wedge, with a background value calculated from the interval $-9 > Z > -11\kpc$ subtracted off (this is to correct for the uniform ``sky'' background). These profiles are only plotted out to $Z=-7.5\kpc$, since at this distance the brightness is only $\sim 1$\% of the background value; beyond this distance the integrated light profiles are too uncertain. Finally, the black profile allows a comparison to the composite bulge plus disk model fitted in Fig.~2, again from a $\pm 10\deg$ minor axis wedge. Since the normalisation between star-counts and surface brightness is unknown, we have shifted vertically the integrated light profiles, as well as the galaxy model profile, to agree with the star-count profile at $Z=-5\kpc$. 

It is patently clear from Fig.~5 that the ACS minor axis star-counts begin to become incomplete at $Z > -4\kpc$,  and get progressively worse as one approaches the galactic plane.\footnote{This was the reason why we chose $Z=-5\kpc$ as the position to normalise the surface brightness profile: further in, the star-counts are incomplete; further out, the uncertainties on the star-count distribution get larger.} Consequently, we ignore the histogram below $|Z| = 4\kpc$, and fit the profile of the ellipsoidal component with a de Vaucouleurs model in the range $-4 > Z > -12\kpc$; this yields an effective radius of $r_e = 1.55\pm0.03$ (the fit is shown with a green dashed line).\footnote{However, over this relatively small range in radius the profile data can also be fit by an exponential function with scale height $h_Z = 2.31\pm0.04\kpc$. Real galaxies can contain significant halo sub-structure \citep[e.g.]{Malin:1979p7053, Ibata:2007p160}, which renders the discussion of the statistical acceptability of different model profiles very uncertain over small radial ranges.} The small difference between this de Vaucouleurs fit to the ``halo'' star-counts and the disk plus bulge model (black line) fit to the global inner galaxy $K$-band light is startling, and suggests an intimate connection between the halo and bulge. Note that the discrepancy between the two curves at $Z > -2\kpc$ is due to the disk model beginning to contribute close to the galactic plane.

\begin{figure*}
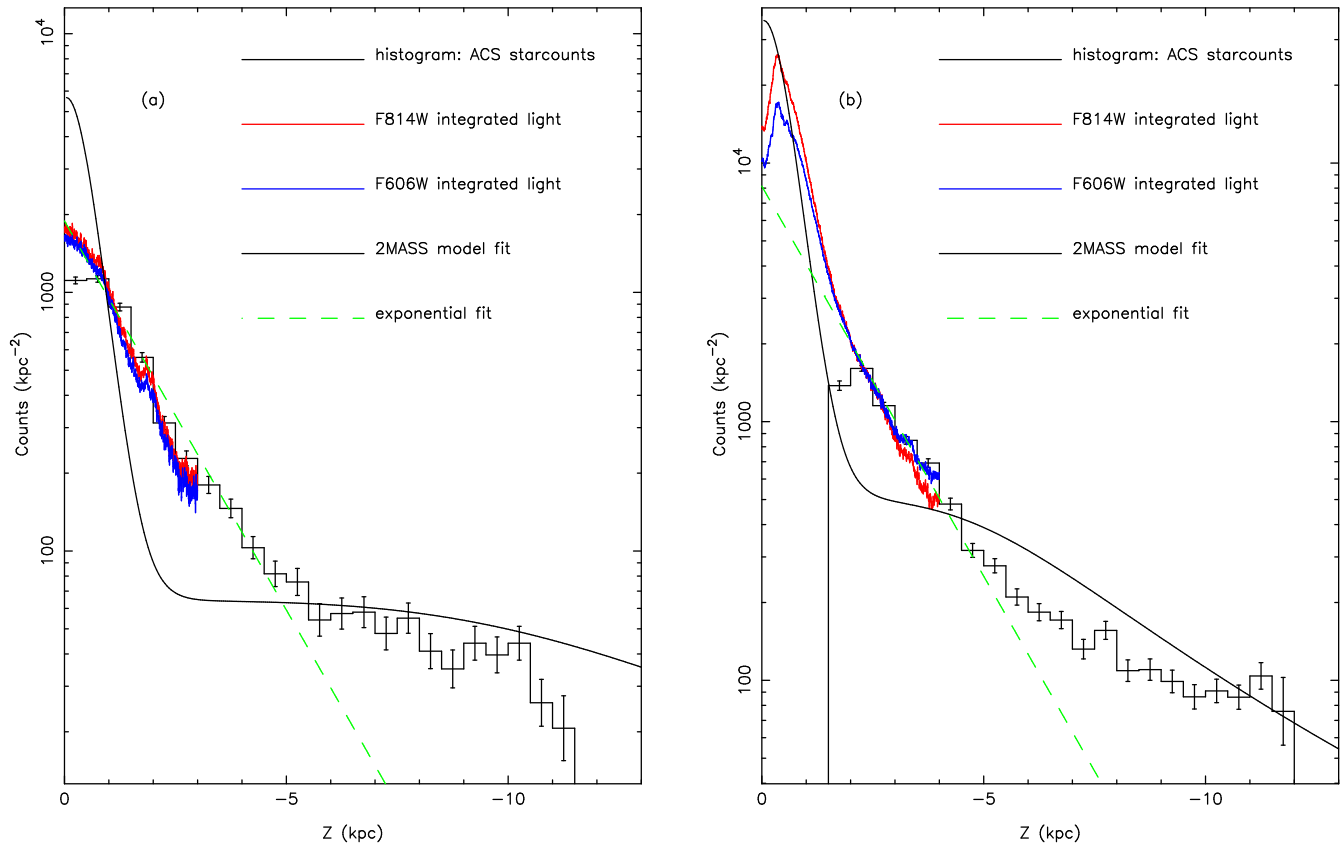

\begin{center}
\hbox{
\hskip -0.25pc
\includegraphics[angle=0,width=20pc]{N891_halo.fig07a.ps}
\hskip 1.75pc
\includegraphics[angle=0,width=20pc]{N891_halo.fig07b.ps}
}
\end{center}
\caption{Panel `a': as Fig.~5, but for a vertical cut at $X = 17\pm 1\kpc$ (the region contained between the green lines in Fig.~3). However, here a fitted exponential model with scale height $h_Z = 1.44\pm0.03\kpc$ is shown instead. The black line shows the two-component galaxy model derived from 2-MASS data. As we discuss in the text, the model fails at $Z > -5\kpc$ because it possesses a thin disk and no thick disk; it is astonishing however, that at $Z < -5\kpc$ the 2MASS bulge model gives a close approximation to the outer halo star-counts. Panel `b': identical to panel `a', except that the vertical cut is made at $X = 8\pm 1\kpc$. The green dashed line is an exponential, constrained to have a scale height $h_Z = 1.44\kpc$, and fit to the starcounts data in the range $-2 > Z > -5\kpc$. Clearly the ``thick disk'' component identified in the outer disk is present also in this region.}
\end{figure*}

The integrated light profiles are also very interesting. As one can see from Figs.~1 and 3, field H1 just misses the centre of the galaxy; this is the reason that the profiles begin at $Z = -0.6\kpc$. Evidently, in the region $-4 > Z > -7.5\kpc$ both the V and I-band light follow each other and the star-counts profile extremely closely. This provides strong evidence that the stellar populations over this region are very similar, and that there is little variation in extinction. We probe this issue further in Fig.~6, where we show, with the black histogram, the distribution of colour difference between the V and I-band profiles. Starting at $Z = -7.5\kpc$ and proceeding towards the galactic plane, the colour differences are extremely small (consistent with zero) until $Z = -5.5\kpc$, where we encounter a large blue dip in the profile (centred at $Z \sim -4.5\kpc$), which clearly cannot be due to extinction. Inspection of the ACS images shows that this feature is likely an artefact due to scattered light (it is a diffuse elongated structure with a relatively sharp outline and no obvious stellar counterpart). Beyond this dip, the profile resumes at its previous level, but then reddens rapidly towards the galactic plane for $Z > -2.5\kpc$. 

Is this reddening of the colour profile towards the galactic plane due to stellar populations differences, or due to extinction? NGC~891 has significant amounts of dust out to $|Z| \sim 1.5$--$2\kpc$ \citep{Howk:1997p7381, Kamphuis:2007p5791}, so internal extinction appears to be a likely culprit. To assess this possibility, we use the \ion{H}{1} maps of \citet{Oosterloo:2007p5583}, and assume a ratio of total hydrogen column density to ${\rm E(B-V)}$ reddening of $(5.94 \pm 0.37) \times 10^{21} \, {\rm atoms \, cm^{-2} \, mag^{-1}}$, identical to that found in the Milky Way \citep{Rachford:2008p8575}. Correcting for {\it half} of this reddening (i.e. assuming that half of the dust is in front of the stars, and half behind) gives the red-lined histogram in Fig.~6, which is consistent with a flat colour profile in the interval $-1 > Z > -3\kpc$. We conclude, therefore, that there is no evidence from the colour profile for a change in stellar populations along the minor axis of NGC~891 between $-1 > Z > -7.5\kpc$.

\subsection{The thick disk}

The other striking feature of Figs.~3 and 4 is the presence of a thick envelope of stars surrounding the disk, reaching down from the Galactic plane to approximately $Z \sim -4\kpc$. This structure is particularly noticeable in field H3 between $15 < X < 19\kpc$. To explore this structure further, we show in panel `a' of Fig.~7 the vertical star-counts profile at $X=17\pm1\kpc$ with a black histogram. The blue and red curves show, respectively, the integrated light profiles in the V and I-bands; as before we have truncated these at the point where they reach 1\% of the background. The good correspondence between the integrated light profiles and the star-counts profile is reassuring, though they highlight the incompleteness problems due to crowding at $|Z| < 0.5\kpc$. Between $-0.5 > Z > -5.0\kpc$ the star counts profile appears exponential; and fitting this region with an exponential model (green dashed line) gives a scale height $h_Z = 1.44\pm0.03\kpc$. This value is very similar to the value of $0.9\pm0.18\kpc$ deduced for the thick disk of the Milky Way \citep{Juric:2008p4886}, and may be precisely such a component. This also will be discussed further in Paper~II. The thin black line shows the profile of the galaxy model fitted to the 2MASS data, and shown previously in Fig.~2, using the same normalisation as in Fig~5. Since the model fit to the 2MASS data does not incorporate a radial cut-off for the thin disk, it vastly over-predicts the counts in the inner $\sim 1\kpc$, and the absence of a thick disk component causes it to under-predict greatly the counts between $-1 > Z > -5\kpc$. However, beyond $Z \sim -5\kpc$ the model predicts a flat star-count distribution, rather close to what is observed. The flatness of the model is partly due to the fact that this vertical cut probes only a relatively limited extent in radius, but also partly because the model is a boxy spheroid. Indeed it is rather remarkable how well that bulge model, fitted primarily to the inner galaxy, works at this great distance.

Panel `b' of Fig.~7 shows an identical analysis for $X = 8\pm 1\kpc$, a location analogous to the Solar radius. Here again one can clearly see the presence of an exponential profile between $-2 > Z > -5\kpc$ sandwiched between the thin disk and the halo. The green dashed line shows an exponential fit to the profile in this radial range, constrained to have a scale height of $h_Z = 1.44\kpc$ (the value measured from the data in panel `a'); this shows nicely that the vertical profile of the thick disk is almost identical between $X=8\kpc$ and $X=17\kpc$. 

\begin{figure}
\begin{center}
\includegraphics[angle=0,width=\hsize]{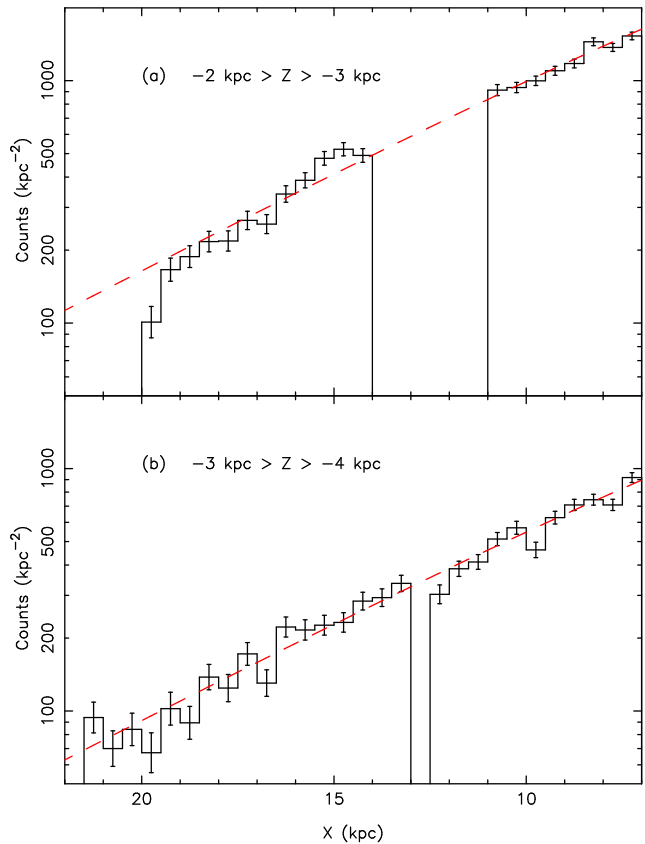}
\end{center}
\caption{Horizontal cuts along the survey, showing the radial fall-off of the thick disk. The top and bottom panels correspond, respectively, to cuts between $-2 > Z > -3\kpc$ and $-3 > Z > -4\kpc$. A projected exponential model is fit to the data in each panel (dashed line), resulting in a best-fit radial scale length of $h_R = 4.81\pm0.11\kpc$ and $h_R = 4.83\pm0.15\kpc$ for the cuts in the top and bottom panels, respectively. The gaps in the data (e.g. at $12.75\kpc$ in panel `b') correspond to bins whose area is more than 20\% missing or masked-out.}
\end{figure}

Fig.~7 shows that we can extract a relatively clean ``thick disk'' sample by selecting stars between $-2 > Z > -4\kpc$: the upper limit is more than 3.5 thin disk scale heights above the galactic plane, while the lower limit lies before the halo-dominated region. The star-counts profile along two horizontal bands in this region is displayed in Fig.~8; the fitted projected exponential model was found to have a radial scale length of $h_R = 4.81\pm0.11\kpc$ in the $-2 > Z > -3\kpc$ band (upper panel) and $h_R = 4.83\pm0.15$ in the $-3 > Z > -4\kpc$ band (lower panel). These values are in excellent agreement with each other and are also similar to the radial scale length value of $h_R=4.19 \pm 0.01\kpc$ derived for the thin disk from the 2MASS $K$-band image in \S2. These values should be refined in future work by performing two-dimensional fits to the disk region. However, this will likely require additional data near the galaxy plane (such data are already present in the archives, but not included in the present study since they are slightly shallower).

\subsection{2-D fit to outer halo}

Armed with the information of the spatial locations where the survey is reliably uncrowded and free of additional components, we can now attempt to fit a two-dimensional function to the star-counts distribution in the halo. To this end, we use again the GALFIT software, fitting only the outer regions of the survey below $Z= -6\kpc$. Although we attempted to fit a general Sersic profile to these starcounts data, we found that the survey does not cover a sufficiently large radial range to fit both the Sersic index and scale length simultaneously. We therefore decided to fix the Sersic index to a value of $n=4$ (i.e. a de Vaucouleurs profile). The resulting best fit (shown in the middle panel of Fig.~4) has scale length $r_e = 1.77\pm0.15\kpc$, flattening of $q = 0.50\pm0.03$ and boxiness of $c = -0.46\pm0.08$. It should be noted that the fit is not good (reduced $\chi^2=1.41$ with 449 degrees of freedom, which has a vanishingly small probability of occurring by chance). The residuals of the fit are displayed on the bottom panel of Fig.~4, and possess considerable non-symmetric large-scale features. Evidently the galaxy halo is not entirely symmetric. As we will see below in \S4 and \S6.4, there also appears to be small-scale spatial clumps in the survey. It is these sub-structures (and to a lesser degree, the large-scale asymmetry), that are responsible for the poor $\chi^2$ of the fit.

\section{Spatial sub-structure}

Having studied the large-scale distribution of stars in NGC~891, we attempt next to search for localised clumps in the number density map of Fig.~4. The necessary first step is to subtract out the smoothly-varying components. We experimented using a two-dimensional Legendre polynomial fit (as performed on the metallicity profile in \S5), but it transpires that the density gradient is too steep to give a satisfactory fit with low-order functions. We therefore returned to the GALFIT software to attempt a 2-dimensional fit of the star-counts distribution. 

\begin{figure}
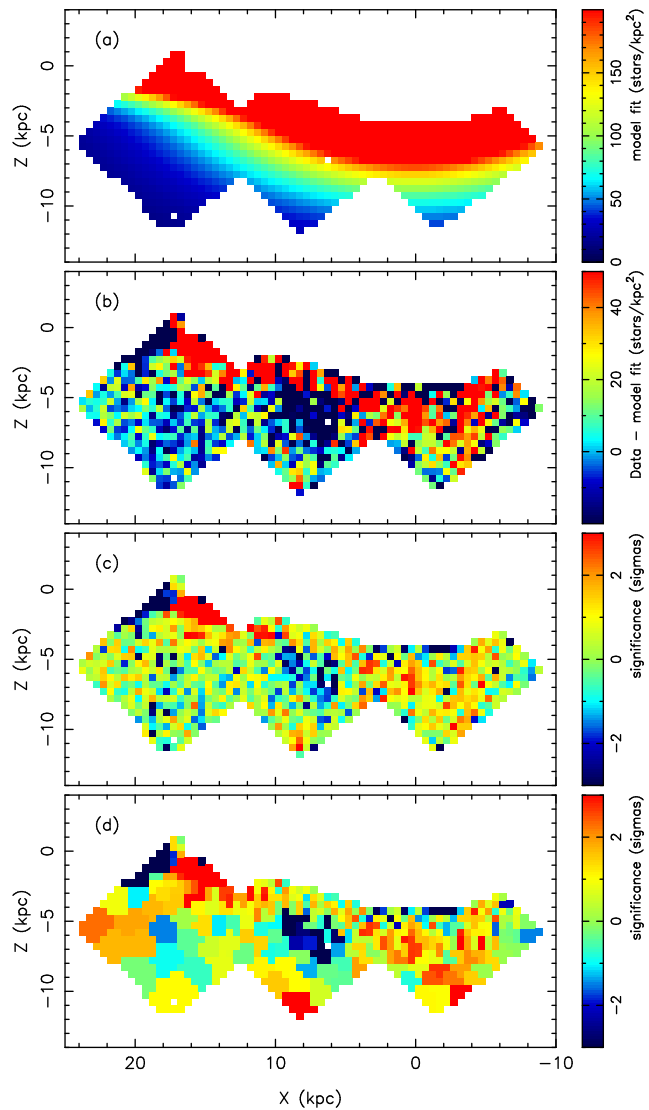

\begin{center}
\includegraphics[angle=0,width=\hsize]{N891_halo.fig09.ps}
\vskip 0.125pc
\includegraphics[angle=0,width=\hsize]{N891_halo.fig09d.ps}
\end{center}
\caption{Panel `a' displays a two-component (disk and spheroid) model fit to the star-counts distribution of Fig.~4; the corresponding residuals are shown in panel `b'. As discussed in the text, the non-truncated disk model cannot fit well the region at $X \sim 18\kpc$, $|Z|<3\kpc$, which results in large scale residuals in this region. A ripple in the residuals centred near $X \sim 5 \kpc$, $Z \sim -7\kpc$ is also apparent. Panel `c' displays the corresponding significance map, revealing several isolated pixels containing significant counts above the model; the same data is reproduced in panel `d', where we have summed over the large Voronoi super-pixels used later for the metallicity analysis in Fig.~11.}
\end{figure}

\begin{figure}
\begin{center}
\includegraphics[angle=0,width=\hsize]{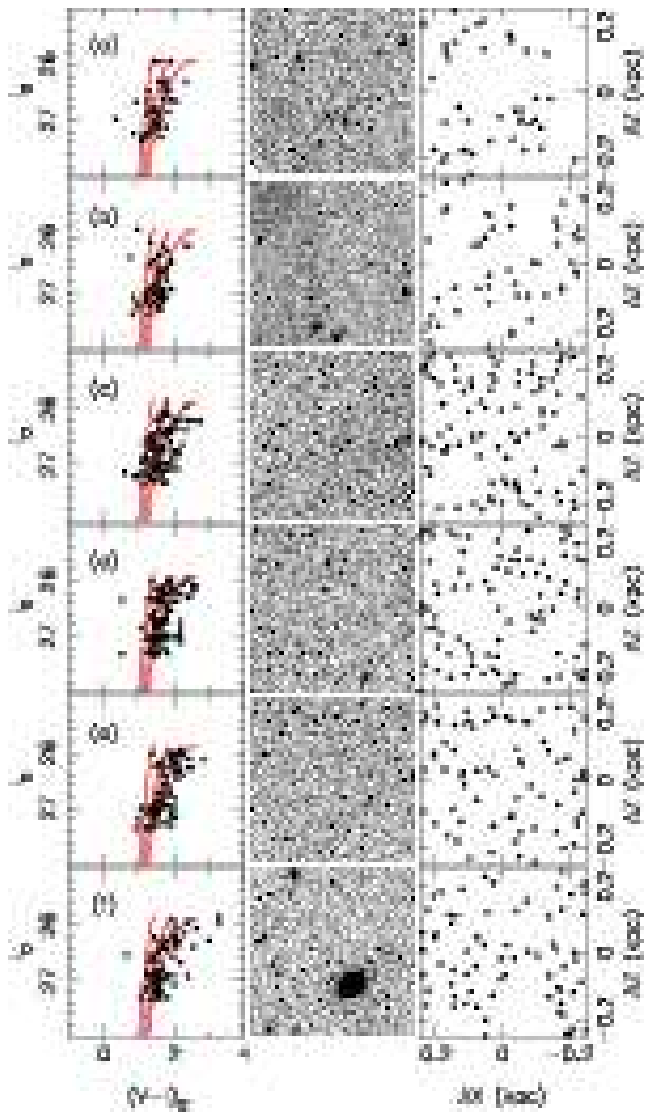}
\end{center}
\caption{The candidate satellites or satellite debris identified via their local deviation from the smooth model shown in Fig.~9. The middle and right-hand panels show the $0.5\kpc\times 0.5\kpc$ pixel where the excess population is detected, while the left-hand panels display the colour-magnitude distribution of the stars in the region. To guide the eye we have overlaid three Galactic globular cluster fiducial sequences, shifted to the distance modulus of NGC~891. These are, from left to right, M~15 (${\rm [Fe/H]=-2.2}$), NGC~1851 (${\rm [Fe/H]=-1.2}$) and 47~Tuc (${\rm [Fe/H]=-0.7}$).}
\end{figure}

To eliminate the effects of incompleteness due to crowding, we masked the data in those regions with a stellar density higher than ${\rm 1000 \, stars \, kpc^{-2}}$ (as suggested by the fits in Figs.~5 and 7). Since the star-counts distribution drops below this value (due to crowding) near the galactic plane, we had to rely on a model to define the boundary in surface density of the mask. However, since the fitted model depends on the boundary, we had to iterate a few times to achieve convergence. A two component galaxy with thick disk and halo was fit to the star-counts. This time, to alleviate the degeneracy of the problem, we fix the centre of both components to $X,Z=(0,0)$, we also fix the scale-length of the thick disk to $h_R=4.83\kpc$ (as found in Fig.~8) and scale-height $h_Z = 1.44\kpc$ (as deduced from Fig.~7a), we further fix the halo to have a de Vaucouleurs profile and we set its boxiness/diskiness parameter to $c=0$ (thus the only free parameters are the central surface brightness of the disk and halo, as well as the halo effective radius $r_e$ and flattening $q$). We stress here that we are not attempting to fit a realistic galaxy model to the star-counts data, but simply trying to obtain a smooth representation of the data. (For completeness, we note that the fitted model had a halo effective radius of $r_e = 1.97\pm0.09\kpc$, and a flattening of $q = 0.55\pm0.01$, which are slightly different values to those fitted previously in \S3.3). 

The residuals between the star-counts distribution (shown previously in Fig.~4) and this fit are displayed in the upper panel of Fig.~9. This residual map (panel `b') shows large-scale structure that highlights the inadequacy of the simple model. In particular, one can see a strong feature at $X \sim 16 \kpc$, $Z \sim -1.5\kpc$ which is due to the modelled disk not being sufficiently dense at that location. The fit cannot accommodate the absence of these stars slightly further out in radial distance (at $X \sim 19 \kpc$, $Z \sim -1.5\kpc$) so compromises between the two. It is possible that this is an artefact of fitting a model with a single exponential component. A further problematic region occurs at ($X \sim 5 \kpc$, $Z \sim -7\kpc$), where one sees a large negative-positive ripple in the residuals (the positive part of the ripple being closer to the plane, at for instance, $X \sim 4 \kpc$, $Z \sim -5\kpc$); curiously, the negative ripple is not reproduced on the symmetric side of the galaxy ($X \sim -5 \kpc$, $Z \sim -7\kpc$).

The significance of the residuals is displayed in the bottom panel, where we have assumed that the only source of uncertainty is the Poisson noise in the star-counts (i.e. that there is no noise in the fitting procedure). Although there are significant peaks around the previously-mentioned large-scale residuals, we also detect several significant isolated spikes. These isolated spikes are candidate satellites of NGC~891 or indeed accretion remnants. In Table~1 we list the positions of the 6 most significant (1-pixel) spikes above $2.5\sigma$ that are well-separated from the areas with problematic residuals; Fig.~10 shows close-up views of the $0.5\kpc\times 0.5\kpc$ super-pixels in which they are found, together with the corresponding colour-magnitude diagrams. Upon inspecting the ACS images visually, candidate satellite `a' is found to possess a plausibly-concentrated distribution (slightly off-centred of the super-pixel centre by $\delta X \sim -0.05$, $\delta Z \sim -0.1$), and a plausibly tight RGB sequence. The nature of the other structures is less clear, although in all cases an RGB sequence is detected.

\begin{figure*}
\begin{center}
\hbox{
\hskip -0.25pc
\includegraphics[angle=0,width=20pc]{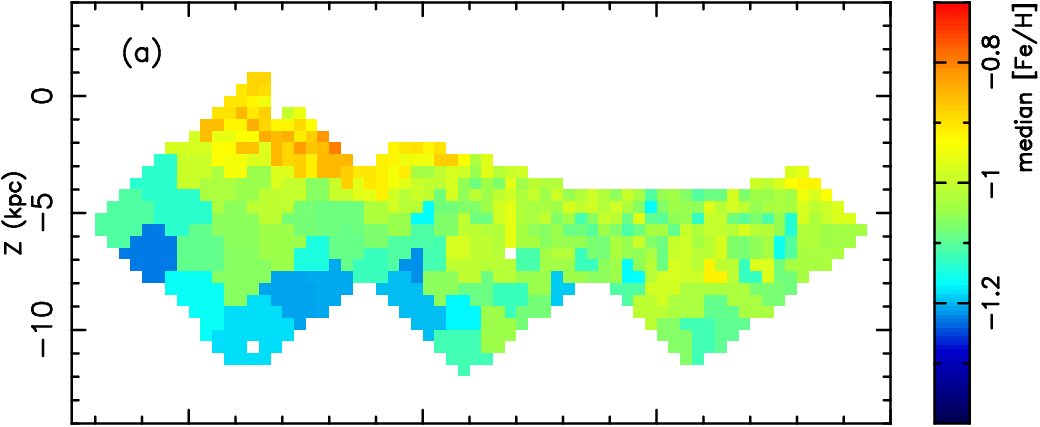}
\hskip 1.75pc
\includegraphics[angle=0,width=20pc]{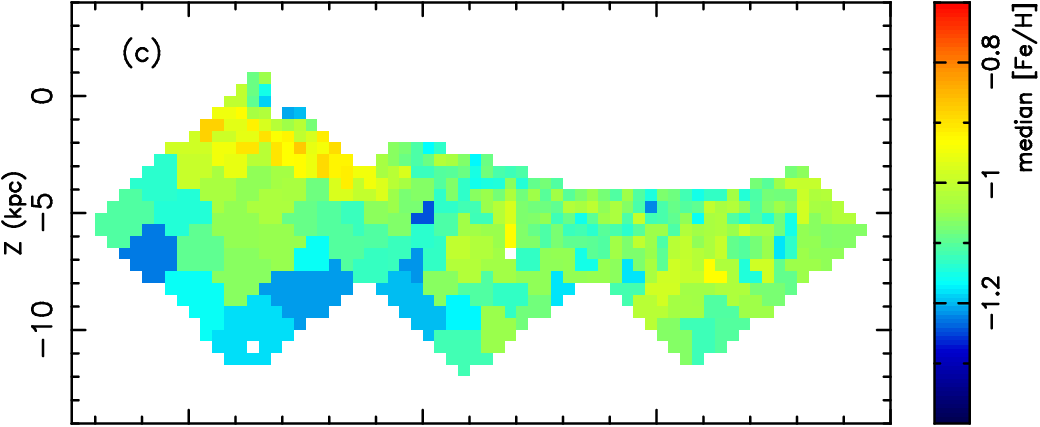}
}
\hbox{
\hskip -0.25pc
\includegraphics[angle=0,width=20pc]{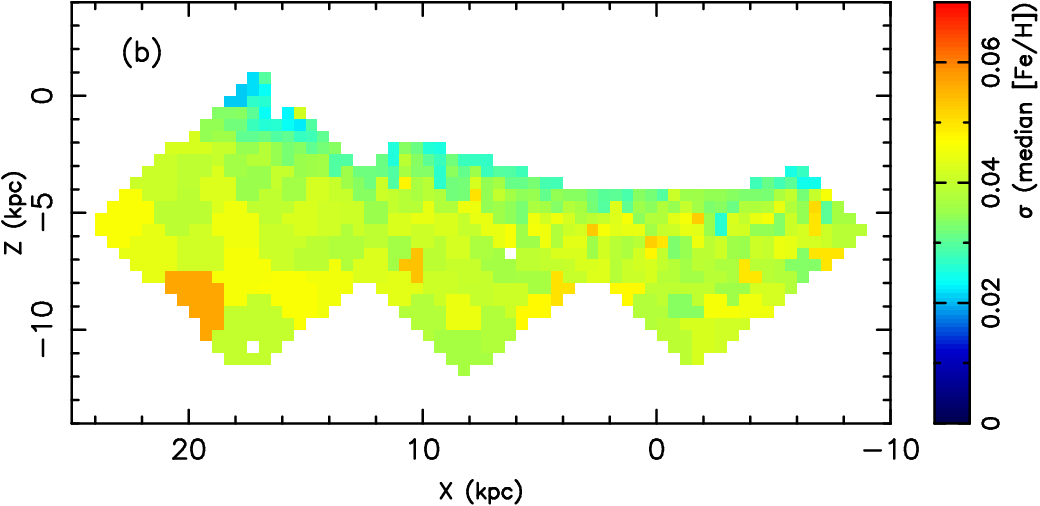}
\hskip 1.75pc
\includegraphics[angle=0,width=20pc]{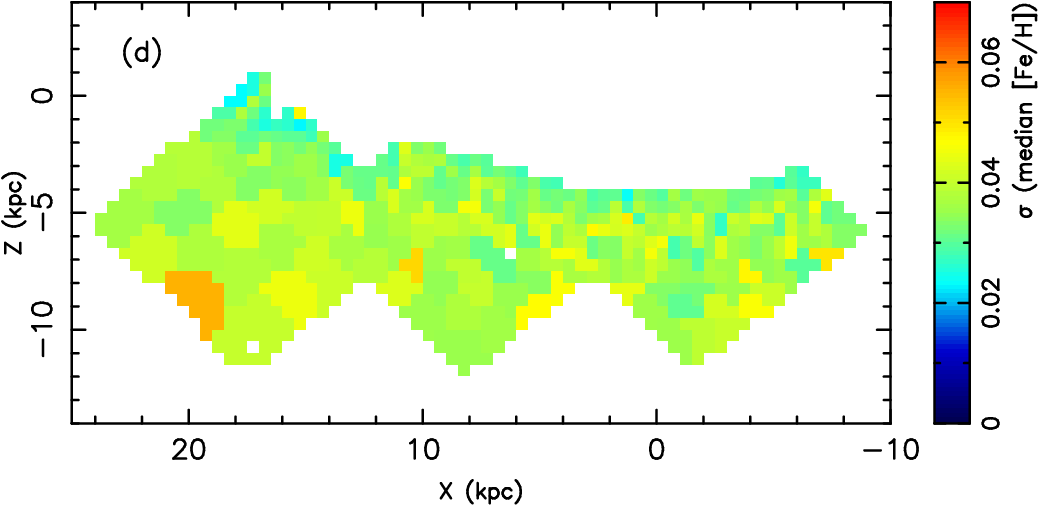}
}
\end{center}
\caption{Panel `a': map of the median metallicity over the ACS survey area, colour-coded with the values shown on the wedge. A large-scale gradient in the halo region is noticeable, as well as a more metal-rich ``thick disk''. Panel `b': map of the uncertainties on the median metallicity, derived from simulating the effect of the photometric measurement uncertainties on the metallicity interpolation. Panels `c' and `d' show the same information as `a' and `b', but with the maximum correction for internal extinction assuming that all the dust associated with the \ion{H}{1} \citep{Oosterloo:2007p5583} is in front of the stars.}
\vskip 1pc
\begin{center}
\hbox{
\hskip -0.25pc
\includegraphics[angle=0,width=20pc]{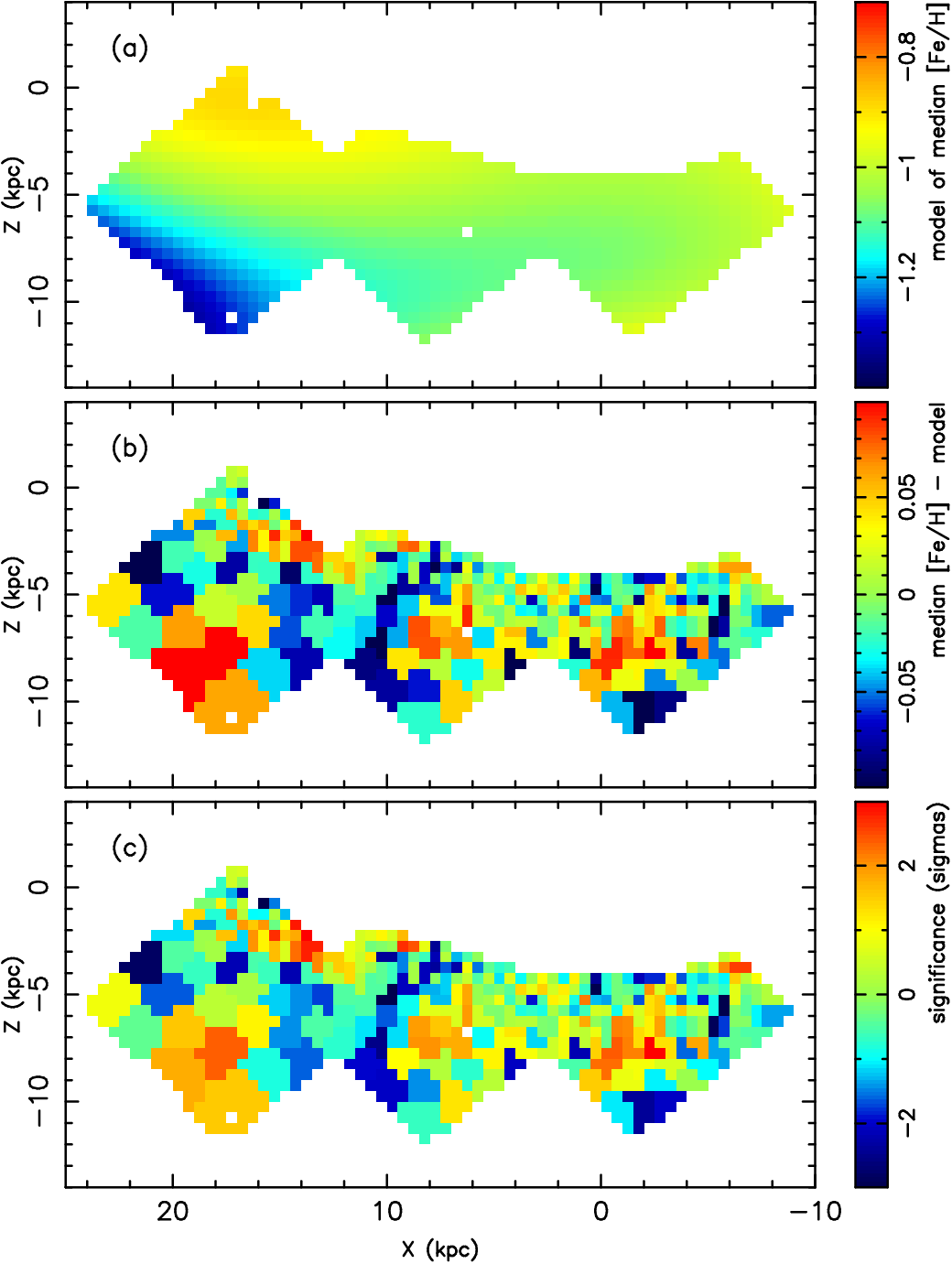}
\hskip 1.75pc
\includegraphics[angle=0,width=20pc]{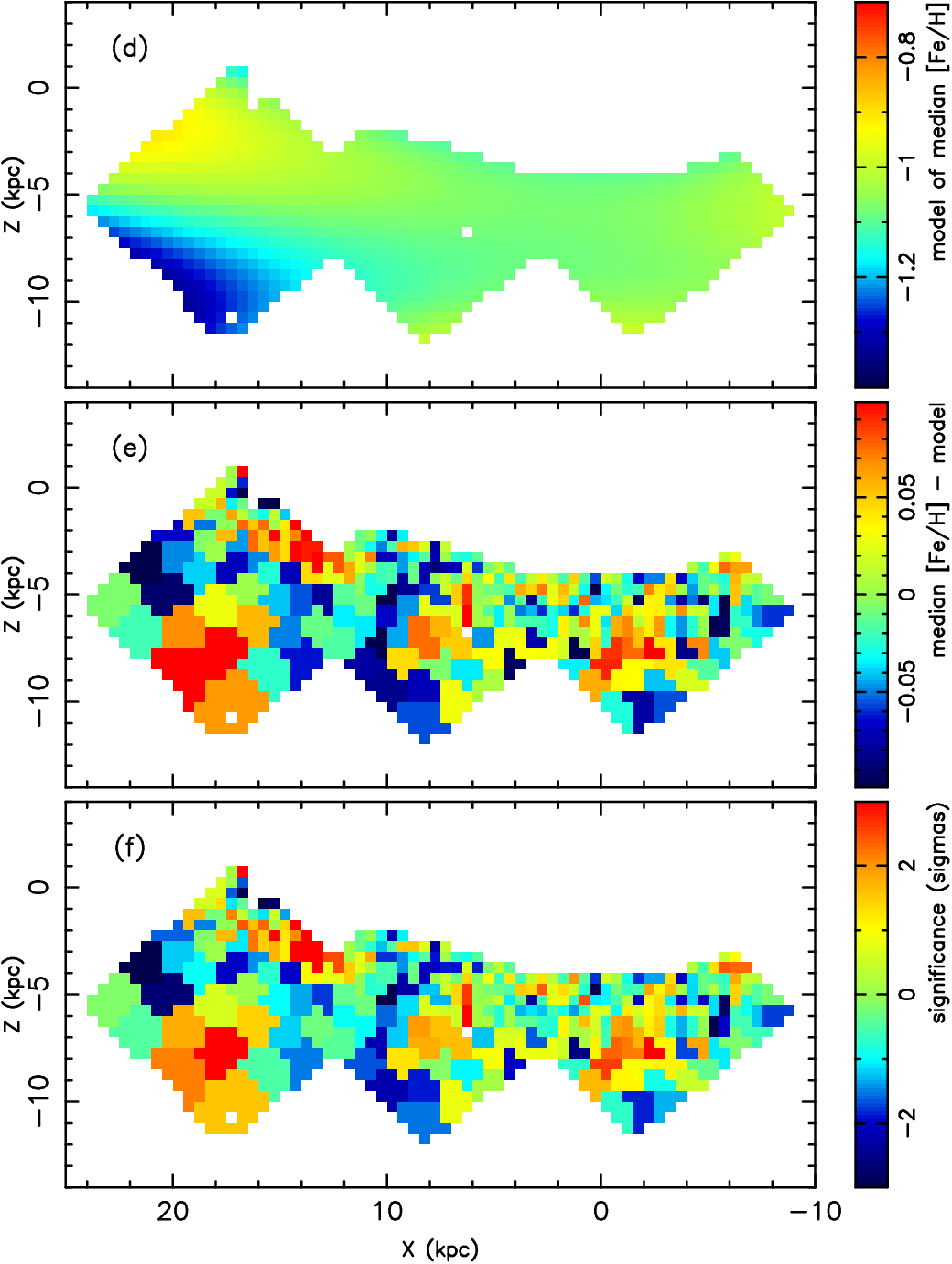}
}
\end{center}
\caption{The inhomogeneities in the median metallicity of the survey stars for the ``standard'' metallicities (left-hand panels) and with the maximum correction for internal extinction (right-hand panels). The upper panels show a Legendre polynomial fit to the median metallicity map. The middle panels display the difference between the median metallicity map and this fit, while the lower panels report the significance level of the differences. These maps show that the halo has highly significant local variations in the median metallicity.}
\end{figure*}

The so-called ``matched filter'' method, where one uses a template stellar population (e.g., an RGB track) and a template spatial distribution (e.g., a 2-dimensional Gaussian) to optimally filter out contaminants, provides a more powerful means to search for such spatial clumps \citep[see, e.g.,][and references therein]{Segall:2007p177}. The analysis is somewhat complicated in the present situation by the strongly varying ``background'' population due to the smooth component of NGC~891, and also due to the complex spatial window function of the ACS survey. Furthermore, the sensitivity varies strongly as a function of the extent of the structure: for instance, one convincing satellite (not listed in Table~1) was detected visually, yet because of its compactness, and hence high stellar crowding, none of its RGB stars are present in the point-source catalogue. We have decided to defer a thorough matched-filter analysis to a subsequent contribution, since quantifying the detection and exclusion limits for the presence of sub-structure will require a lengthy discussion beyond the scope of the present contribution.

\begin{table}
\caption{Candidate sub-structures.}
\label{tab1}
\begin{tabular}{crrc}
\hline
\hline
Label in Fig.~10 & $X (\kpc)$  & $Z (\kpc)$ & Significance (sigmas) \\
\hline
a &   8.25 &-11.25 &  3.6\\
b &  -2.75 & -9.75 &  3.1\\
c &  0.75 & -7.25 &  3.1\\
d &  -4.75 & -6.75 &  3.0\\
e &  11.25 & -4.75 &  2.5\\
f &  16.25 & -3.25 &  3.1\\
\hline
\hline
\end{tabular}
\end{table}

\section{Spatial distribution of the median metallicity}

We now proceed to map out the metallicity properties of the surveyed stars. The metallicity of each star passing the quality tests laid out in \S3 (with the additional constraints ${\rm I_0< 27.0}$, and DOLPHOT uncertainties ${\rm \delta I < 0.1}$ and ${\rm \delta V < 0.15}$),  was derived by interpolating between the models of \citet{VandenBerg:2006p4971} of $\alpha$-enhanced red giant branch stars of mass $0.8\msun$, as discussed in Paper~I. The models span ${\rm -2.314 < [Fe/H] < -0.397}$; all stars outside of the colour-magnitude range covered by these RGB tracks were flagged and not used in subsequent analysis (i.e., no extrapolation beyond the validity of the models was attempted). 

After initially binning the survey area into $0.5\kpc\times0.5\kpc$ pixels, we employed the Voronoi tessellation algorithm of \citet{Cappellari:2003p7062} to implement a set of super-pixels in which the Poisson noise level of the star-counts exceeded $S/N=10$. As we have discussed in Paper~II, the RGB stars in all the galactic components identified in NGC~891 cover a wide range in colour, implying a vast metallicity range. This intrinsic spread of stellar populations makes the differences from population to population rather difficult to highlight, so we decided to make use of the median metallicity as our population statistic, as this is one of the most robust measures that can be made of a distribution. This has the additional advantage that it is not affected in any way by the spatial density of the population, neither is it affected by issues such as inter-CCD gaps, holes due to bright stars etc, so in many ways it is one of the most sensitive and robust measures that one can make regarding the spatial properties of the galaxy. The map of the median value of the metallicity calculated in the Voronoi super-pixels is displayed in panel `a' of Fig.~11. 

There are two main sources of uncertainty in calculating the median metallicity map: a measurement uncertainty arising from the photon noise in the magnitude measurements of the stars, and a sampling uncertainty due to the limited number of stars per bin. Ideally, we would resample from the true metallicity distribution in each pixel, using the estimated photometric uncertainties to quantify the effects of photon noise. While this would give the correct total uncertainty, it is not possible to implement, since the true ${\rm [Fe/H]}$ distribution in each pixel is unknown.

We therefore proceed as follows to estimate the errors. To this end, we adopt the photometric measurement uncertainties returned by DOLPHOT; these estimates are reliable, as we show in Paper~II by comparing the photometry of stars detected in the overlapping region of fields H2 and H3.
\footnote{We have also undertaken all of the following analysis adopting the alternative choice of using the mean magnitude-uncertainty relations derived in Paper~II. The difference between the two methods of estimating the uncertainty is small and the conclusions of this study remain identical. We chose to adopt the DOLPHOT error, because we judged that it was more reliable to have a star-by-star uncertainty estimate, especially for the outer halo region where crowding incompleteness is not an issue.}

\begin{itemize}
\item
{\it Step 1:} The photometry of each of the $N_i$ stars in super-pixel $i$ is randomly drawn from a Gaussian distribution of dispersion equal to the measurement uncertainty (and with a mean equal to the actual measurement), and its metallicity estimated by interpolation from the \citet{VandenBerg:2006p4971} tracks in exactly the same manner as for the survey stars. This procedure is repeated 1000 times for each super-pixel. The resulting dispersion in the median values calculated from these Monte-Carlo simulations $\tilde{\sigma}_{i, \rm real}$, using random realisations of the real stars actually present in each bin, gives a good estimate of the effect of the photometric measurement errors, but the sampling uncertainties have not yet been properly taken into account (the tilde is meant to highlight this fact).

As mentioned above, we have no means of knowing the underlying ${\rm [Fe/H]}$ distribution in each super-pixel. However, we can calculate how much the Monte-Carlo procedure in ``step 1'' underestimates the total uncertainty, if the underlying distribution were the observed distribution over the whole survey. 

\item
{\it Step 2:} To assess this, we draw an initial set of $N_i$ stars at random from the full data-set, and then, using the corresponding photometric uncertainties, draw 1000 random realisations from this initial set, in an identical way to ``step 1''. The dispersion of the median metallicity value  $\tilde{\sigma}_{i, \rm fake}$, is calculated from this sample. The whole procedure is repeated 1000 times, in order to derive the average dispersion of the median $\langle \tilde{\sigma}_{i, \rm fake} \rangle$, that would be expected from applying the procedure outlined in ``step 1'' to a super-pixel with $N_i$ stars if there were no spatial stellar population gradients.

\item
{\it Step 3:} Finally, we draw $N_i$ stars at random from the full data-set (again with the photometry drawn randomly from a Gaussian distribution of dispersion equal to the corresponding photometric measurement uncertainties), and derive the median metallicity. The dispersion of the derived value over 1000 simulations gives the typical total uncertainty $\sigma_{i, \rm fake}$ in the median ${\rm [Fe/H]}$ in super-pixel $i$ with $N_i$ stars, which again is only strictly valid if the metallicity distribution does not depend on spatial position.
\end{itemize}

The difference between the result of ``step 3'' and ``step 2'' gives the correction that needs to be applied to ``step 1'' to account properly for the sampling noise. Thus the final uncertainty on the median metallicity of bin $i$ is calculated as $\sigma_i = \tilde{\sigma}_{i, \rm real} + \sigma_{i, \rm fake} - \langle \tilde{\sigma}_{i, \rm fake} \rangle$, where the last two terms are the corrections to the procedure outlined in ``step 1''. We note that these terms are corrections to an uncertainty, and therefore are not added in quadrature. The estimated uncertainties $\sigma_i$ are displayed in panel `b' of Fig.~11.

As discussed in \S3, there is evidence for substantial amounts of dust close to the plane of the galaxy in NGC~891. In an attempt to assess and correct the effects of the dust, we again use the \ion{H}{1} map of \citet{Oosterloo:2007p5583} to estimate the $E(B-V)$ reddening following the Galactic calibration of \citet{Rachford:2008p8575}. In the following analysis, we will present the metallicity distributions with and without this additional extinction correction. Note that this is actually a maximum correction, assuming that all the extinction lies in front of the stars. The corresponding internal extinction-corrected median metallicity map and its uncertainty map are shown in panels `c' and `d' of Fig.~11. Although the correction is substantial near the galactic plane (as can be seen in the corrected colour profile of Fig.~6), the correction declines rapidly away from the plane, so that at the outer edge of the ACS survey region, the estimated internal extinction as derived in this way from the \ion{H}{1} column density amounts to less than 0.001~mag. 

\begin{figure}
\begin{center}
\includegraphics[angle=-90,width=\hsize]{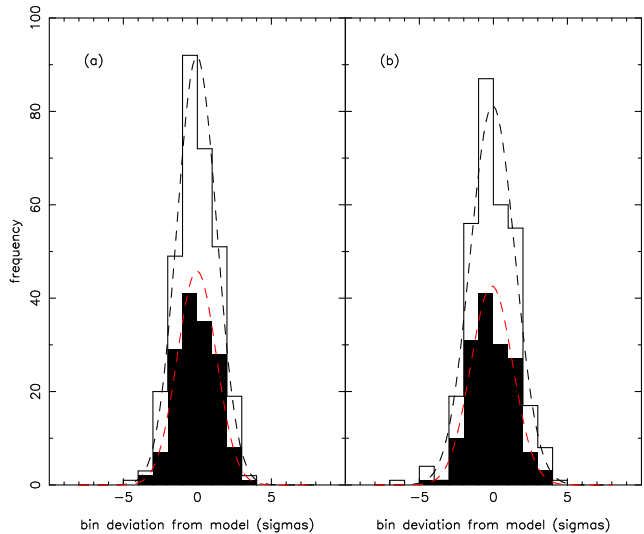}
\end{center}
\caption{Panel a: the distribution of the statistical significance of the deviations from the model in panel `c' of Fig.~12, showing a large number of highly significant super-pixels. The sample has a vanishingly small probability of being drawn by chance from a Gaussian distribution of unit dispersion. The filled-in histogram shows the pixels in the pure halo region below $Z=-5\kpc$. Panel `b' shows the same information for the sample that was corrected for internal extinction (panel `f' of Fig.~12). This again cannot occur by chance. The dashed lines show a fitted Gaussian of dispersion $\sigma=1.4$ (1.3) in the left-hand panel and $\sigma=1.5$ (1.4) in the right-hand panel, where the values in brackets refer to the halo-only sample.}
\end{figure}

Fig.~11 (both panels `a' and `c') reveal a large-scale radial gradient in the median metallicity of the halo, consistent with the results presented in Paper~I; this can be seen clearly by inspecting the gradient along the $X$-direction at $Z \sim -10\kpc$. We perceive also that the median metallicity of the outer ``thick disk'' component is significantly more metal-rich than that of the halo, consistent with expectations. We delay further discussion of the behaviour of the metallicity gradient to \S6.

The issue we wish to draw attention to here is the fact that the uncertainties on the median metallicities in the super-pixels are small, typically $\sim 0.04$~dex, compared to the large pixel to pixel variations. This can be seen explicitly in Fig.~12; here the left-hand panels show the distributions not corrected for internal extinction, with the right-hand panels showing the internal extinction-corrected versions. Since we do not have an a-priori model of the large-scale distribution of metallicity, we adopt a pragmatic approach of fitting the observed distributions in Fig.~11 (panels `a' and `c') with a two-dimensional Legendre polynomial (up to order 3 in both $X$ and $Z$, i.e. 10 parameters). The difference between the distribution of median metallicity and this ``model'' is displayed in the middle panels of Fig.~12, while the bottom panels show the significance level of these differences. An inspection of the lower two panels reveals that the super-pixels have very significant localised variations in median metallicity that occur on the scale of a super-pixel or a few super-pixels. The actual distribution of the statistical significance of the variations in panels `c' and `f' of Fig.~12 is displayed in Fig.~13, which demonstrates clearly that the variations are much larger than what one would expect from a halo with a smooth spatial distribution in median metallicity. 

We tested the statistical reliability of our analysis method by ``scrambling'' the photometry of the survey, i.e. keeping the positional information and super-pixel positions intact, but randomly reassigning the photometric measurements of the stars. This was undertaken only for those stars at $Z < -5\kpc$, which can be considered to represent a halo-only sample, that way we avoid mixing in more metal-rich populations from close to the galactic plane. The underlying smooth distribution was refit for each random realisation. Reassuringly, the resulting map of differences between data and model for this halo-only sample has a distribution of significance levels consistent with having been drawn from a Gaussian of unit dispersion with better than 10\% probability.

\section{Discussion}

\begin{figure}
\begin{center}
\includegraphics[angle=0,width=\hsize]{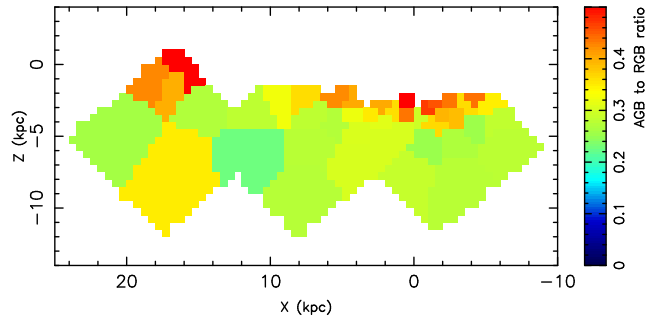}
\end{center}
\caption{Map of the AGB to RGB star ratio. The AGB star candidates are confined to regions close to the disk of the galaxy, indicating that the halo beyond $|Z|>4\kpc$ shows no evidence for intermediate age stellar populations.}
\end{figure}

\subsection{Median metallicity maps}

The median metallicity map (Fig.~11a), which shows both large-scale metallicity gradients and small-scale metallicity clumps, is one of the major results of this survey. However, before interpreting these maps as indicative of metallicity variations, it is important to consider the alternative: that they are revealing differences in mean age. Due to the well-known age-metallicity degeneracy, a redder mean colour in a pixel can either correspond to a more metal-rich population or else an older population. One avenue to examine this possibility is to examine the ratio of the spatial distribution of asymptotic giant branch (AGB) and RGB stars. Bright AGB stars trace predominantly intermediate age ($2-6\Gyr$) populations \citep{Mouhcine:2002p9435}, whereas RGB stars will predominantly represent more ancient populations. Their ratio therefore gives a relative map of the number of intermediate to old stars in the galaxy. 

We selected candidate AGB stars to be those point-sources with $24.34 < I_0 < 25.84$ with all other quality criteria identical to the RGB sample discussed in \S4. As before, we amalgamated the $0.5\kpc\times0.5\kpc$ pixels into super-pixels using the Voronoi tessellation algorithm of \citet{Cappellari:2003p7062}, using a target signal-to-noise ratio of 10; the resulting map is shown in Fig.~14. There are substantially fewer AGB star candidates than RGB stars, so the algorithm is forced to generate very large super-pixels, especially in regions of low AGB density. The map shows immediately that the AGB star candidates are predominantly confined close to the disk of the galaxy. The completeness tests presented in Paper~II show that at $Z < -2\kpc$ the stellar distributions should not be significantly contaminated by blends. We therefore conclude that, since the intermediate age AGB stars appear not to be present in the halo regions, there is no evidence for variations in population age beyond $Z < -4\kpc$, and we therefore more confidently interpret colour variations as variations in metallicity.

The colour scale in Fig.~11a was chosen to allow direct comparison to the median metallicity map for Galactic stars in the SDSS \citep[][their Fig.~18]{Ivezic:2008p4821}. This analysis of the SDSS covers much of the radial range $3 < R < 16\kpc$, and vertical range $0.5 < |Z| < 8\kpc$ in the Galaxy; and of course probes the Solar neighbourhood disk. In contrast, the present ACS star-counts survey is hopelessly incomplete in the thin disk, however it extends out much further in radius and vertical height. Essentially all of the metallicity gradient in the Milky Way occurs at $|Z| < 4\kpc$, which is only very poorly covered in the ACS fields H1 and H2. Nevertheless, we detect in fields H2 and H3 a clear vertical gradient between $Z\sim -3\kpc$ and $Z\sim -5 \kpc$, rather similar to what is seen in the Milky Way, albeit offset $\sim 0.3$~dex more metal-rich. 

We detect a significant drop in metallicity at large radius, such that the outermost regions of field H3 are approximately $\sim -0.2$~dex more metal-poor than the inner halo at $Z=-5\kpc$ on the minor axis (see also the discussion in Paper~I). In contrast, the region of the Galactic halo covered in the analysis of \citet{Ivezic:2008p4821} is mostly uniform, with a slight positive metallicity gradient, due to the presence of the Monoceros ring structure at the upper radial extent of the surveyed region. 

The effect of the Monoceros Ring on the Galactic metallicity distribution cautions against a simple interpretation of the differences in the halo components of the two galaxies. It is still possible that both NGC~891 and the Milky Way possess similar underlying ancient halos, that are radially well-mixed (over the scales surveyed) and metal-poor. By analogy to the Milky Way, the slightly more metal-rich inner halo ($r \simlt 15\kpc$) of NGC~891 may simply be due to a Monoceros Ring-like accretion event which is now better spatially mixed than the Monoceros Ring. However, if the stellar halo of NGC 891 is indeed dominated by intermediate metal-rich stars, the implications are profound: a moderately metal-rich stellar halo suggests an origin in the disruption of chemically evolved accreted fragments, which in turn has an implication on the mass function of small mass dark matter halos. Determining the dominant mechanisms that produce such a variation between galaxies of similar mass is a key problem for models of spiral galaxy formation \citep{Renda:2005p187, Font:2008p2999}.

The SDSS analysis presented by \citet{Carollo:2007p8649} suggests that the Galactic halo is composed of two distinct populations, an inner flattened halo with prograde rotation with a metallicity distribution that peaks at ${\rm [Fe/H] \sim -1.5}$ and an outer more spherical component, with retrograde rotation that peaks at ${\rm [Fe/H] \sim -2.0}$ and that begins to dominate at radii between approximately 15 and $20\kpc$. Inspection of the Legendre polynomial fits to the median metallicity in Fig.~12 (panels `a' and `d') reveals that this inner/outer halo dichotomy may be present in NGC~891 as well. In particular, the Fig.~12d, where we applied the correction for internal extinction, shows that the halo metallicity is approximately constant everywhere except the outer edge of field H3 where the metallicity drops suddenly (at $r \sim 15\kpc$). Only a more spatially extensive survey will tell whether this reflects the global properties of a putative outer halo, or even whether the median metallicity continues to drop with distance to attain values consistent with those measured in the Galaxy.

\begin{figure*}
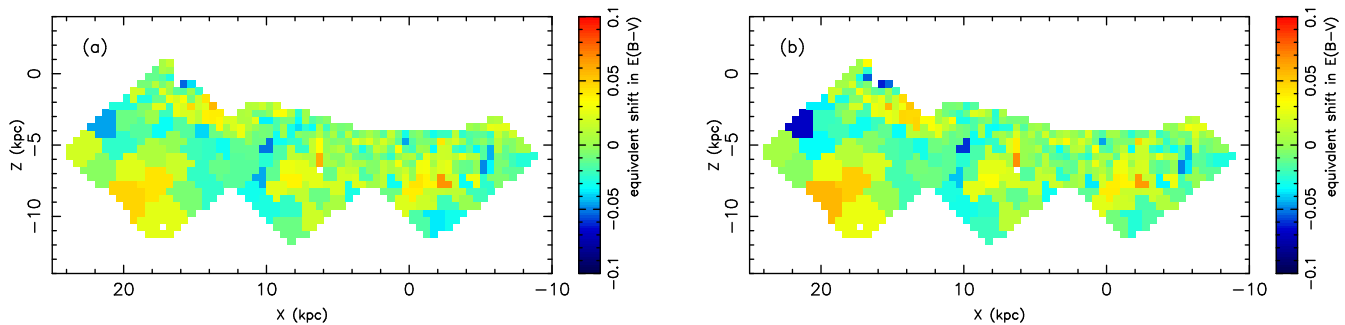

\begin{center}
\hbox{
\hskip -0.25pc
\includegraphics[angle=0,width=20pc]{N891_halo.fig15a.ps}
\hskip 1.75pc
\includegraphics[angle=0,width=20pc]{N891_halo.fig15b.ps}
}
\end{center}
\caption{The shift in reddening, compared to the adopted mean value of 0.065 from \citet{Schlegel:1998p5331}, required to compensate for the apparent chemical inhomogeneities.  Panel `b' shows the same information for the sample that was corrected for internal extinction}
\end{figure*}

\subsection{Metallicity clumps}

The analysis presented in \S4 indicates that in addition to the large-scale variations discussed above, the median metallicity varies significantly on the smallest scales that we can probe with the present dataset, $\sim 1\kpc$. This result is highly unexpected, especially for the halo component, since we observe the galaxy in projection, and are integrating the population over a large range along the line of sight. It is important therefore to consider all the possible explanations for this observation.

\subsubsection{Photometric zero-point?}
A first possibility is that the photometric zero-point is not constant over the ACS fields. This could be caused by, for instance, substantial flat-fielding errors. Simulations showed that, on average in the RGB selection region we employed, $d {\rm [Fe/H]} / d (V-I) = 1.08$, so that very substantial variations in the zero-point (of order $0.1$~mags) are required to induce the observed clumpiness in metallicity. The ACS is an extremely stable and well-calibrated instrument, and is routinely used to derive exquisite photometry ($\delta (V-I) < 0.01$) of globular clusters which cover the field of view of the instrument \citep[see, e.g.][]{Sarajedini:2007p7496, Piotto:2007p7359}. Clearly the ACS does not generally suffer from such flat-fielding errors. Furthermore, the observed metallicity clumping pattern in the present survey was not reproduced from field to field. This possibility can be eliminated.

\subsubsection{Photometric errors underestimated?}
A second possibility is that the photometric errors were poorly estimated. In this case the metallicity uncertainties would be correspondingly larger, diminishing the significance of the detections. We have run simulations to investigate this possibility, and find that this can indeed cause the observed affect, though the uncertainties would have to have been underestimated by a factor of $\sim 3$. However, the photometry tests in the overlap regions presented in Paper~II provide a very robust and reliable calibration of the uncertainties, and are consistent with the artificial star tests; we find it highly unlikely that the photometry errors were significantly underestimated, and a factor of 3 is completely out of the question.  We judge therefore that this possibility can be also be safely eliminated.

\subsubsection{Foreground and background contamination?}

As we have discussed in Paper~I, contamination from foreground or background sources is negligible. Based on the Besan{\c c}on Galaxy synthesis model \citep{Robin:2003p2969}, we expect $\sim 119$ stars in the range $24.5 < I_0 < 27$ in the entire survey area. The number of background sources is also very limited, as demonstrated by the Hubble Deep Fields North and South which contain, respectively, 10 and 22 point-like sources (FWHM $ < 0\scnd 2$) with magnitudes $22 < F814W < 26$ \citep{Casertano:2000p7095}. The ACS survey of NGC~891 should therefore be essentially free of such contaminants. There should be, on average, less than one contaminant per Voronoi super-pixel in Fig.~11, and therefore foreground and background contaminants should have no significant effect on the median metallicity values.

\subsubsection{Foreground dust?}
It is also possible that we are detecting the effects of clumpiness in the foreground Galactic inter-stellar medium. Though probing other lines of sight, photometric studies of globular clusters with the Hubble Space Telescope, which show tight colour-magnitude sequences (especially at the main-sequence turnoff) again constrain the scatter around the mean extinction. In the large sample of globular clusters studied by \citet{Sarajedini:2007p7496}, the typical scatter over an ACS field of view was $\sigma_{\rm E(B-V)} \sim 0.01$~mag. In fields of higher extinction such as that of the globular cluster NGC~2808, where the mean reddening is ${\rm E(B-V)=0.23}$, accurate photometry constrains the differential reddening to $0.02$~mag \citep{Bedin:2000p7362, Piotto:2007p7359}. As mentioned above, the mean reddening towards NGC~891 is ${\rm E(B-V)=0.065}$ \citep{Schlegel:1998p5331}, so under the reasonable assumption that reddening variations scale approximately linearly with mean reddening, we expect less than one hundredth of a magnitude scatter in ${\rm E(B-V)}$ from foreground dust. 

In Fig.~15 we probe the shift in extinction, compared to the mean value of 0.065, that would be required to offset the deduced difference in metallicity. For this exercise we simulated adding or subtracting extinction to each Voronoi super-pixel, re-computing the metallicity measurements on a star-by-star basis (extinction coefficients of $A_V/E(B-V)=3.315$ and $A_I/E(B-V)=1.940$ were assumed). Although the rms dispersion of the map is only $\sigma({\rm E(B-V)})=0.025$~mag, it can be seen that some of the required offsets are huge, and are particularly problematic in metal-poor pixels, which occasionally would require impossible negative extinction. For this reason, this possibility also appears unlikely. Note that the \citet{Schlegel:1998p5331} maps show relatively low variations in extinction towards NGC~891, as can be seen in Fig.~16.

\subsubsection{Dust in NGC~891 itself?} 

NGC~891 is well-known to exhibit extra-planar dust features out to surprisingly large distances; indeed some of these structures can even be seen directly in the upper panel of Fig.~3. The vertical structures appear to be related to galactic fountain or chimney phenomena, and driven by supernovae. Could dust in NGC~891 itself be causing the observed variations in colour? Dust has been detected at $|Z| < 1.5\kpc$ \citep{Howk:1997p7381}, and molecular gas is also detected over a very similar vertical range \citep{GarciaBurillo:1992p7443}. However, these distances are of course very close to the galactic plane in the context of the present study.

In contrast, \ion{H}{1} gas has recently been detected over essentially the entire vertical range of the present ACS survey \citep{Oosterloo:2007p5583}, forming a halo-like envelope around the galaxy. Earlier \ion{H}{1} observations  suggested that the outer gaseous disk was highly flared \citep{Sancisi:1979p11158} or extremely warped into the line of sight with the near side in the hemisphere probed by our survey \citep{Becquaert:1997p7459}. However, subsequent modelling of the full velocity cube \citep{Swaters:1997p7803} contradicted these conclusions, favouring a non-warped configuration. The very deep observations of \citet{Oosterloo:2007p5583}, which are among the deepest ever reached in any galaxy, show filaments and counter-rotating \ion{H}{1} structures, similar to those observed around M31 \citep{Braun:2004p2943}. A significant fraction of this gas is likely to be of accretion origin.

The maps of \citet{Oosterloo:2007p5583} show that on the minor axis between $Z=-5\kpc$ and $Z=-12\kpc$, the \ion{H}{1} column density drops from $10^{20} {\rm cm}^{-2}$ to $10^{19} {\rm cm}^{-2}$. If we assume the same ratio of hydrogen column density to extinction as in the Milky Way \citep{Rachford:2008p8575}, this corresponds to a mean extinction of 0.017 to 0.0017~mag. However, given the contribution from accretions, the halo gas in NGC~891 is probably less abundant in dust than the Milky Way at the Solar neighbourhood, so the reddening in the halo of NGC~891 should be even lower than these estimates suggest. We therefore consider that dust in the target galaxy itself is unlikely to have contributed significantly to the observed small-scale variations in median colour beyond $Z < -5\kpc$.

\begin{figure}
\begin{center}
\includegraphics[angle=-90,width=\hsize]{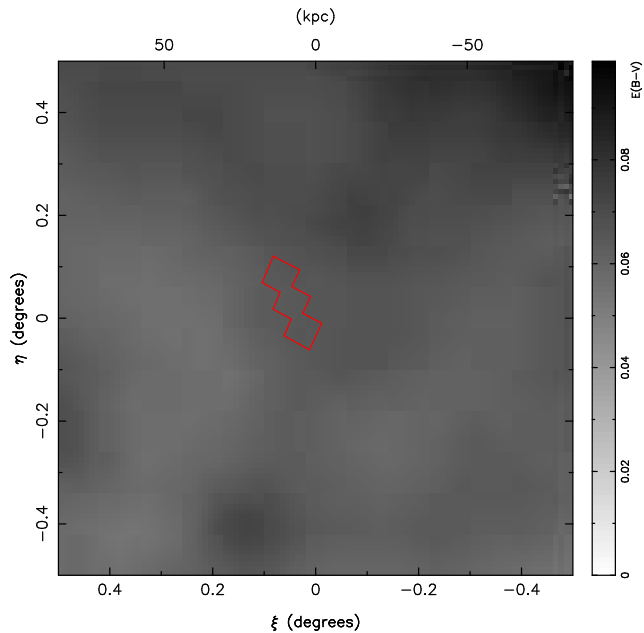}
\end{center}
\caption{Map of the reddening in the vicinity of NGC~891 (gnomonic projection centred on the galaxy), derived from the \citet{Schlegel:1998p5331} dust maps. The variations in the mean extinction are clearly modest over this 1 sq. degree region, although the map obviously does not have the resolution required to probe the reddening variations over the ACS survey area (the marked polygon), much less to evaluate whether the small spatial scale variations suggested by Fig.~15 are realistic or not.}
\end{figure}

\subsection{Implications of the metallicity variations}

The remaining choice is then that the observed colour variations imply real variations in metallicity. This is a fascinating possibility and would suggest that the ACS survey has revealed the existence of numerous accretion remnants that are spread over a large volume in the halo and are still far from being fully phase mixed. These remnants, which are identified by the change that they cause in the median metallicity of the average halo population, must nevertheless enhance the local density of the halo over the average value at the particular location of the structure. Why then (with the possible exception of the stream-like structure between $X \sim 1\kpc$, $Z \sim -9\kpc$ and $X \sim -4\kpc$, $Z \sim -6\kpc$) is there no obvious correspondence between density variations (Fig.~9d) and metallicity variations (Fig.~11a)? This may be explained by the fact that it is {\it much} harder to identify density sub-structures in the presence of a strong density gradient (the surface density changes by more than an order of magnitude just in the halo region). The gradient itself will diminish the contrast of the structure, and errors in the model subtraction are likely to be much larger in a steep distribution. In comparison, the large-scale metallicity gradient (Fig.~12a) is very gentle, which of course makes the survey considerably more sensitive for detecting localised variations in the median metallicity. 

The number of stars that are needed to cause the apparent metallicity variations turns out to be rather modest. Of course the actual fraction will depend on the metallicity distribution of the interloping population, but taking a somewhat extreme case of a single stellar population of uniform metallicity ${\rm [Fe/H]=-0.7}$ (like 47~Tuc), we find that in a typical Voronoi super-pixel a -0.1~dex metallicity variation in Fig.~12b can be caused by a additional population numbering only $\sim 15$\% of the putative ``normal'' smooth halo. If this is indeed the case, it is then not surprising that the population cannot be detected from its density contrast, since this is smaller than the change in density over a typical super-pixel.

It is worth noting that due to the virtually negligible contamination from foreground or background sources, the present observations are extremely sensitive to extended sub-structures. It is in principle possible with the ACS to detect a population of $\sim 15$ RGB stars scattered over a volume of many hundred cubic kpc. Such sensitivity is orders of magnitude greater than what can currently be achieved in the Milky Way with the SDSS (although the SDSS performs extremely well at detecting concentrated structures, and has access to distance information). This would explain why sub-structures analogous to those apparent in Fig.~12 are so hard to detect directly in the Galactic halo.

The detection of such a highly (sub-)structured halo such as suggested by Figs.~12c and 12f has profound implications for galaxy formation models. Hierarchical formation scenarios predict an abundance of sub-structures such as streams and merger remnants (see, e.g. \citealt{Bullock:2005p2996}). Indeed, the Via Lactea~II simulation \citep{Diemand:2008p8589}, the highest resolution simulation of the evolution of dark matter in a Milky Way like galaxy that has been published to date contains $\sim 40000$ dark matter satellites down to $10^6\msun$, some of which are tidally disrupted to form streams, especially in the inner regions of the modelled galaxy. It is possible that the chemical in-homogeneities revealed in Fig.~12 are caused by the disrupted remnants of this vast predicted population. 

Observationally, it is now well-established that the Galactic halo contains numerous signatures of accretion \citep{Belokurov:2006p7995}, some of which, such as the giant stream of the Sagittarius dwarf, are still on-going events \citep{Ibata:1994p417, Ibata:2001p369, Ibata:2001p333}. Likewise, our giant neighbour Andromeda also displays a multitude of streams, shells, spurs and satellites \citep{Ibata:2001p334, Ibata:2007p160, Ferguson:2002p300}. However, what is surprising about the results presented here on NGC~891 is that the sub-structure appears ubiquitous: in the halo at $Z < -5\kpc$, 52\% of the Voronoi super-pixels in Fig.~12 deviate by more than $1\sigma$ from the smooth model.  

Although this appears astonishing, \citet{Bell:2008p6111} came to a similar conclusion in the Milky Way. These authors analysed the statistics of density deviations from a smooth model of the halo star-count distribution in the SDSS. They accomplished this by measuring the following statistic from a sample of halo stars:
\begin{equation}
\sigma/{\rm total} = { {\sqrt{ {{1}\over{n}} \sum_i (D_i - M_i)^2 - {{1}\over{n}} \sum_i (M_i^\prime - M_i)^2}} \over {{{1}\over{n}} \sum_i D_i}} \, ,
\end{equation}
where $n$ is the number of discrete bins the sample is divided into, $D_i$ are the individual observed counts in those bins, $M_i$ are the model values in the pixels, and $M_i^\prime$ is a Poisson realisation of the model with mean $M_i$. Thus the nominator is the pixel scatter of the data around the model minus the expected scatter in the model, while the denominator was chosen to be the total of the Poisson scatter. This statistic is convenient because it is independent of the choice of binning scale, and the Poisson noise contribution is removed. \citet{Bell:2008p6111} found that that their statistic exceeds $\sigma/{\rm total} = 0.4$ in the Galactic halo region probed by the SDSS, and furthermore that projecting the stream-only models of \citet{Bullock:2005p2996} (i.e., those models with no underlying smooth halo component), and analysing them in the same way as the data, gave similarly large values of $\sigma/{\rm total}$. This indicates that the Galactic halo is substantially lumpy and also that it is approximately as lumpy as predicted by the most extreme hierarchical formation models. 

The $\sigma/{\rm total}$ statistic is related to the star-counts density, and so not directly applicable to our median metallicity maps. However, we can easily estimate the scatter that the $\sigma/{\rm total} = 0.4$ value corresponds to in a bin of 100~stars (the target density for Voronoi super-pixels in the halo region in Figs.~11 and 12). Solving Eqn.~1 for $M_i$ with $D_i=100$ and $n=1$, gives model values of $M_i=59$ or $M_i=142$. We would therefore have interpreted the detection as a $(D_i-M_i) / \sqrt{D_i} = 4.1\sigma$ peak or equivalently a $-4.2\sigma$ dip. This average scatter in density is substantially larger than the metallicity scatter that we report here, though this should be expected. First, our findings result from integrating along the line of sight through the entire halo, which must necessarily reduce the contrast of any sub-structure, whereas  \citet{Bell:2008p6111} are able to break the halo into a (small) number of shells in distance. Second, the sub-structures will almost certainly have overlapping metallicity distributions, which again must reduce the contrast. Thus we find that the present results on the variations of the median metallicity in NGC~891 are qualitatively similar to those of \citet{Bell:2008p6111}, and hence support hierarchical halo formation scenarios, although further work is now needed to compare our observations to simulations, including the effect of chemical enrichment in halo sub-structures. 

It has not escaped our notice that the clumpy halo suggested by the metallicity variations we have just been discussing appears inconsistent with the de Vaucouleurs profile that we have fit to the halo component of the galaxy in \S2 and \S3. The de Vaucouleurs profile is known to be found in remnants of major mergers \citep[see, e.g.,][and references therein]{Naab:2006p8590}, and is associated with structurally well-mixed populations. While this may be the case for the  smooth halo component, it is nevertheless possible that there are numerous additional small accretion structures that arrived after the main event that caused the formation of the bulk of the halo. We therefore judge that this apparent inconsistency is not too worrying, unless the majority of the halo is composed of sub-structure (but see the next section below).

\subsection{Spatial clumpiness}

In the analysis in \S4 and Fig.~9 we performed a simple model fit to the spatial distribution of RGB stars, which revealed certain systematic residuals as well as several candidate dwarf satellite galaxies. The strong systematic residuals render useless most of the area of the map for the purpose of quantifying the incidence of low-level sub-structure, in the manner we have just undertaken in \S6.3 with the distribution of median metallicity. However, the region at $X>10\kpc$ and $Z<-4\kpc$ appears devoid of nasty residuals or any very significant localised peaks in Fig.~9c. For this region we calculate $\sigma/{\rm total} = 0.14$, lower than the value obtained by \citet{Bell:2008p6111} for the Milky Way. The probability of finding such a large value of this statistic by chance is 0.8\%, assuming the smooth model of Fig.~9a. As we have mentioned above, we view a projection of NGC~891, whereas the \citet{Bell:2008p6111} resolved the Milky Way into various shells in distance. Thus one would expect a weaker sub-structure signal in NGC~891, consistent with what is seen.

It should be noted that the ACS survey is very different to the SDSS, with very different photometric problems. For instance, as far as the \citet{Bell:2008p6111} analysis is concerned, one could worry about the photometric calibration of the SDSS over the sky, the effect of foreground and background contamination, the consequences of the strategy of observing in long ``stripes'', the effect of corrections for dust over large fractions of the sky, etc. The present ACS study has a different set of  problems. The fact that we measure similar density fluctuations in both galaxies is therefore very reassuring for both analyses.

\subsection{Large-scale structures}

Despite the detection of a strong signal due to small-scale clumping, the spatial distribution of extra-planar stars in NGC~891 shows (Fig.~9) a surprisingly smooth large-scale distribution over a physical scale where the distribution of stars in M31 is highly clumpy with loops, spurs and streams of relatively high surface brightness ($\mu_V \sim 27$--$28 {\rm mag/arcsec^2}$) \citep{Ibata:2001p334, Ferguson:2002p300}. It would appear therefore that the main stellar accretion events that NGC~891 has experienced were either not as massive as those experienced by Andromeda, or that they happened longer ago and are now more smeared out. With the rather limited spatial coverage currently available for NGC~891, however, it is difficult to push this comparison much further.

\subsection{What is the relationship between the halo and bulge?}

In \S2 we saw that fitting a two component model to the 2MASS data at $|Z| \simgt 0.5\kpc$ (the limit is approximate because the mask in panel `b' of Fig.~2 is irregular) gives a good fit with a bulge model that has a Sersic index of $n=4.33\pm0.1$, that is, a model that is essentially a de Vaucouleurs profile. Further out on the minor axis at $Z < -4\kpc$, a region which one would consider a-priori to be halo-dominated, both the integrated light profile and the star counts follow this same profile, and even the outer halo (\S3.3) is consistent with the same $r^{1/4}$-law. This begs the question whether there is a structural difference between the bulge and the halo.

Unfortunately, the current data sets do not constrain the surface brightness profile of the inner bulge region satisfactorily. As argued by \citet{Balcells:2003p7070}, high spatial resolution data is required to measure accurately the Sersic index of bulges: seeing effects in ground-based surveys smear out point-sources in the nucleus of the galaxy, leading to erroneous high Sersic index values. They find that bulges have Sersic indices $\overline{n} = 1.7 \pm0.7$. Thus despite its relative immunity to extinction, the 2MASS $K$-band data (with its $\sim 2\arcsec$ resolution) is not particularly useful to probe the inner bulge profile. At the time of writing, the Hubble Space Telescope archive has a single high-resolution NICMOS/NIC3 image in F160W (i.e. approximately H-band) that covers the very centre of the galaxy. However, even this NICMOS image suffers significantly from the strong interstellar extinction that blankets the centre of NGC~891, rendering any study of the central light profile highly uncertain.

In light of the \citet{Balcells:2003p7070} study, the paucity of reliable data in the inner galaxy forces us to interpret our findings cautiously. The boxy  bulge in NGC~891 seen clearly in Fig.~2, has most probably in reality a Sersic profile with a low (exponential-like) index, consistent with the modern picture of these structures as having been formed from disk instabilities \citep[see, e.g.][]{Athanassoula:2008p7204}. The fact that we derive de Vaucouleurs-like index values in the fit to the $K$-band 2MASS data is likely due to the fact that we have included data out to $|Z| < 2.35$ where the halo component begins to dominate. The reasonable match of this same model to the outer halo profile (at $Z < -5\kpc$ in Fig.~7a) indicates that this de Vaucouleurs-like profile is a reasonable model for the entire {\it halo} region probed in the ACS survey. The ellipsoidal structure detected in panel `a' of Fig.~4, and fitted with a ``disky'' elliptical structure in \S3.3, therefore need not have a direct relationship with the inner boxy bulge. Nevertheless, it will be interesting to obtain further high-resolution infra-red imaging of  the centre of this galaxy to actually measure the Sersic index of the bulge.

\subsection{Thick disk}

Thick disks are now known to be a ubiquitous component of galaxies \citep{Yoachim:2006p8530}. These structures may hold important clues to the early formation of disk galaxies, as they may be in themselves  fossil evidence of the formation mechanism. Currently it is not clear whether these structures form by secular processes that redistribute stars  in the galaxy \citep[see, e.g.,][]{Kroupa:2002p9148, Roskar:2008p9305}, whether they are the result of the heating of the early thin disk via a significant merger \citep{Quinn:1993p9322}, or else due to repeated accretions of low-mass satellites over the lifetime of the galaxy \citep{Abadi:2003p8664}.

NGC~891 is an excellent target to study the thick disk component: it is edge-on and close enough to resolve individual stars, and far enough to afford us a panoramic view. The thick disk of NGC~891 was first detected in integrated light by \citet{vanderKruit:1981p5356}, and later confirmed by \citet{Morrison:1997p8565}, who found that this component possesses a scale height between 1.5--$2.5\kpc$, and comprises between 3\% and 21\% of the surface brightness of the thin disk at the galactic plane. \citet{Morrison:1997p8565} conclude that the thick and thin disks have approximately twice the scale height of the Milky Way disks, which seems to be consistent with the larger extent of the \ion{H}{1} and molecular gas in this galaxy perpendicular to the disk. The thick disk structure has also been detected from the mid-infrared emission of asymptotic giant branch stars, which possess a scale-height of $1.3\pm 0.3\kpc$ \citep{Burgdorf:2007p5738}. However, the present contribution provides the first detection of the thick disk in this galaxy from star-counts, which gives much higher signal to noise for fitting purposes. It is interesting that the thick disk component is structurally very close to a perfect exponential disk, with a scale height $h_Z = 1.44\pm0.03\kpc$ at $X=17\kpc$ which remains essentially constant to $X=8\kpc$, and with a scale length of $h_R = 4.8 \pm 0.1 \kpc$, larger, but rather close to the value of $4.19\pm 0.01 \kpc$ deduced for the thin disk. 
\footnote{We have refrained from re-analysing the relative masses of the thick and thin disk, since new ACS data cover the disk at $X \sim 12\kpc$, and we feel that this question can be addressed much better with that additional field.}
For comparison, the Galactic thick disk has a scale length of $h_R = 3.6\kpc$, substantially longer than the $h_R = 2.6\kpc$ scale length of the thin disk, implying a substantial scatter between galaxies in the relative properties of the two disk components. The well-behaved exponential implies a well-settled component in dynamical equilibrium. We defer a discussion of the implications of this finding to Paper~II, where we will additionally investigate the stellar populations in the component as well as their profile. 

In a sample of edge-on galaxies studied in integrated light, \citet{Yoachim:2006p8530} find the following scaling relationships between radial scale length and circular velocity $V_c$: 
$h_{R,thin} = 3.4 (V_c/100\kms)^{1.2} \kpc$, $h_{R,thick} = 3.9 (V_c/100\kms)^{1.0} \kpc \, .$
Given the measured circular velocity of $V_c=224\kms$ \citep{Paturel:2003p9422}, these relations predict $h_{R,thin} = 8.95\kpc$ and $h_{R,thick} = 8.74$, completely inconsistent with our measurements in NGC~891. It is curious that these relations also do not work for the Milky Way; it will be interesting to investigate why these two galaxies, which after all are among the best cases one can study, seem to deviate so markedly from the global trend found by \citet{Yoachim:2006p8530}.

\section{Conclusions}

We have undertaken a structural and colour (i.e. chemical) analysis of a large survey of resolved stars in the edge-on galaxy NGC~891. Extending out to $X \sim 24\kpc$ and $Z\sim -12\kpc$, the survey provides a unique view of the stellar populations in the outer bulge, halo and thick disk of a galaxy whose optical properties are very similar to those of the Milky Way.

Our main findings can be summarised as follows:
\begin{enumerate}
\item
We detect the thick disk of NGC~891 in star-counts, and give the first measurement of the global structural properties of the component based on these measurements, which are inherently much more reliable than earlier measurements based on integrated light. We find that in the spatial window of this survey, the thick disk follows closely an exponential disk profile with scale length of $h_R=4.8\pm0.1\kpc$. The scale height is found to be $h_Z=1.44\pm0.03\kpc$. The smooth exponential nature of the component suggests that it is now dynamically well mixed. We will discuss the stellar population properties of the component further in Paper~II.
\item
The large-scale metallicity distribution is approximately uniform in the halo component out to $\sim 15\kpc$, with a median value of ${\rm [Fe/H] \sim -1.1}$. Furthermore, the colour of this inner region of the halo is identical to that of the bulge beyond $\sim 0.5\kpc$ (after accounting for internal reddening due to dust in NGC~891 itself). At a radial distance of approximately $15\kpc$ the halo metallicity drops to ${\rm [Fe/H] \sim -1.3}$, and in analogy to the Milky Way \citep{Carollo:2007p8649}, this is suggestive of the presence of a distinct outer halo component. Further data beyond the present survey limits are required to confirm this possibility.
\item
Along the minor axis, the structural profile of the halo merges smoothly into that of the bulge, with the combination of the two following closely a de Vaucouleurs profile from $\sim 0.5\kpc$ out to the edge of the survey. However, the shape of the distribution clearly changes with distance, being rounder ($q=0.73$) close to the bulge, and substantially flatter ($q=0.50$) at large distance. The similarity of the bulge and halo in both colour and structural properties is striking, though we reserve judgement on the relationship between the two due to the difficulty of studying the structure of the bulge region inside $0.5\kpc$.
\item
The most important finding of this work is the discovery of small-scale sub-structure in the median metallicity maps (Figs.~11 and 12). Though the metallicity variations are small, they are nevertheless highly significant, and appear to indicate that the halo of NGC~891 is composed of a multitude of small accretions that have not yet been fully blended into a ``normal'' smooth halo. This result based on metallicity variations, is in broad agreement with the findings of \citet{Bell:2008p6111}, based on density variations in the Milky Way. We also repeat the analysis of \citet{Bell:2008p6111} on the stellar density distribution in the halo, and find that their $\sigma/{\rm total}$ statistic, the fractional rms deviation of the data from a smooth model, is  $\sigma/{\rm total} = 0.14$. This is somewhat lower than the value of $0.4$ found in the Milky Way, but may be consistent with it given that we view NGC~891 in projection.
\item
We detect directly 6 low-mass satellite candidates from their local density enhancements; these are possible dwarf spheroidal galaxies, or tidally disrupted fragments thereof. However, further work is needed to implement a more sophisticated detection algorithm (such as a matched filter), to quantify the detection limits and physical properties of the satellites, and to compare these to hierarchical formation scenarios of low-mass structures in a $\Lambda$-CDM cosmology.
\end{enumerate}

Thus we have found that the halo of NGC~891 shows evidence for substantial amounts of sub-structure. These clumps show similar statistics in the scatter of stellar density as the Milky Way, a finding which supports the most extreme models of hierarchical halo formation. We have also detected the presence of these sub-structures from the small-scale statistical fluctuations in the metallicity distribution, the first-time it has been possible in any galaxy, and which gives complementary information on the nature of the sub-structures. Substantial follow-up work is now required to undertake cosmological galaxy formation simulations, incorporating star-formation and chemical evolution in the dark matter mini-halos and to compare the resulting stellar distributions to our observations. 

The current ACS survey is unfortunately rather limited in spatial extent and many of the questions left open by this study will require a more panoramic view to be addressed. It will be interesting also to extend this analysis to other extra-galactic systems, and in particular to assess the level of sub-structure as a function of galaxy morphology, type, mass, and accretion history.

\section*{Acknowledgements}

RI would like to thank Chien Peng warmly for updating his GALFIT package to include a projected exponential disk model, and making the latest beta version of the software available to us. We would also like to thank Tom Oosterloo for providing us with his very deep \ion{H}{1} map of the galaxy, and Caroline Bot, Bernd Vollmer and Fran{\c c}oise Combes  for helpful discussions. We acknowledge the usage of the HyperLeda database (http://leda.univ-lyon1.fr). This research has made use of the SIMBAD database, operated at CDS, Strasbourg, France.

\bibliography{N891_halo.bib}
\bibliographystyle{mn2e}

\end{document}